%% file: main.tex
\documentclass[%
	journal=jctcce,
	manuscript=article
]{achemso}

\usepackage[utf8]{inputenc}
\usepackage{amsmath}
\usepackage{bm}
\usepackage{pgf} 

\newcommand{\figref}[1]{Fig.~\ref{#1}}
\newcommand{\bra}[1]{\left \langle #1 \right |}
\newcommand{\ket}[1]{\left | #1 \right \rangle}
\newcommand{\braket}[2]{\left \langle #1 \right | \left . #2 \right \rangle}
\newcommand{\Tr}[1]{\text{Tr}\left ( #1 \right )}
\DeclareMathOperator{\trace}{Tr}
\newcommand{\ci}{\mathrm{i}}
\newcommand{\eu}{\mathrm{e}}

\newcommand{\revisiondiff}{0}
\newcommand{\revisionnew}{1}

\newcommand{\revisiontype}{\revisiondiff}
\newcommand{\revision}[2]{%
\if\revisiontype\revisiondiff
     {\color{red}{#1}}{\color{green}#2}
\else
     \if\revisiontype\revisionnew
         #2
     \else
         #1
     \fi
\fi}

\author{Štěpán Marek}
\email{stepan.marek@physik.uni-regensburg.de}
\author{Jan Wilhelm}
\affiliation{Regensburg Center for Ultrafast Nanoscopy and Institute of Theoretical Physics,
University of Regensburg,
Regensburg, Germany}

\title[RT-BSE Implementation]{Linear and Nonlinear Optical Properties of Molecules from Real-Time Propagation Based on the Bethe-Salpeter Equation}

\abbreviations{BSE,ETRS,BCH,TDDFT,SHG}

\begin{document}


\begin{abstract}
We present a real-time propagation method for computing linear and nonlinear optical properties of molecules based on the Bethe-Salpeter equation.
The method follows the time evolution of the one-particle density matrix under an external electric field.
We include electron-electron interaction effects through a self-energy based on the screened exchange approximation.
Quasiparticle energies are taken from a prior \textit{GW} calculation to construct the effective single-particle Hamiltonian and we represent all operators and wavefunctions in an atom-centered Gaussian basis.
We benchmark the accuracy of the real-time propagation against the standard linear-response Bethe-Salpeter equation using a set of organic molecules.
We find very good agreement when computing linear-response isotropic polarizability spectra from both approaches, with a mean absolute deviation of 30~meV in peak positions.
Beyond linear response, we simulate second harmonic generation and optical rectification in a non-centrosymmetric molecule.
%
%
We foresee broad applicability of  real-time propagation based on the Bethe-Salpeter equation for the study of linear and nonlinear optical properties of molecules as the method has a similar computational cost as time-dependent density functional theory with hybrid functionals. 
\end{abstract}

\section{Introduction}

Recent advances in the optical microscopy\cite{NOTESidayHuber,TEPLYangHou},
scanning tunneling microscopy\cite{LWSTMCockerHuber} and
laser pulse control\cite{AttosecondCorkumKrausz,AttosecondKrauszIvanov}
have significantly increased the amount of experimentally accessible information
about the light-driven excitations of matter. In many investigations,
non-linear optical response\cite{NonlinearHerrmanWilhelmSoavi,HHGSchmidHuber,HHGSTVHernandezGarcia,HHGMoleculesItataniCorkum} properties
become important.
This calls for theory of non-linear optical responses which could help to gain further insights
into the mechanism behind the observed nonlinear optical phenomena. 

Commonly used methods for the theoretical description of optical excitations and absorption spectra
from first principles include time-dependent density functional theory\cite{ElectronicStructureMartin,TDDFTUllrich,TDDFTLi,RTTDDFTHeHe,RTTDDFTAttaccaliteGruning}
(TDDFT) and the $GW$-Bethe-Salpeter equation ($GW$-BSE) approach\cite{ElectronicStructureMartin,ExcitationsReviewOnidaRubio,SiliconClustersRohlfingLouie}.
However, these methods are subject to certain limitations.
For TDDFT, when using local and semilocal approximate exchange-correlation potentials, 
the excitonic effects, stemming from the screened Coulomb interaction between the excited charge carriers, are
described inacurrately or may be missing entirely\cite{TDDFTUllrich}. This limitation in TDDFT can be improved by using more advanced functionals, such as tuned hybrid functionals\cite{ElectronicStructureMartin}. However, developing a single functional that reliably describes excited-state phenomena across a broad range of matter, including metals, atomically thin semiconductors, and molecules, remains a major challenge.
Furthermore, the computational
cost of hybrid functionals is significantly higher than the cost of local or semilocal functionals.

Compared to TDDFT with local or semilocal functionals, the $GW$–BSE approach provides a more accurate description of
excitonic effects and optical spectra, as $GW$-BSE naturally includes the screened Coulomb interaction between the excited charge carriers\cite{ExcitationsReviewOnidaRubio,BSEBlaseLoos}.
Common implementations of $GW$-BSE~\cite{BSELiuBlum,TurbomoleBSEHolzerKlopper,MolGWBrunevalNeaton,BSEBlaseLoos,ADFBSEFoersterVisscher,BSEToelleNeugebauer,YamboSangalliMarini,GWBSEZhouKanai}  employ the Casida framework to study  electronic excitations in the linear response limit, so for relatively weak driving electric fields. 
Due to advances in laser spectroscopy~\cite{Boyd2008}, also nonlinear optical excitations are commonly studied, which go beyond this Casida framework.  
Specific nonlinear optical properties, in particular low-order responsenses in second and third order, can be computed in frequency domain in $GW$-BSE~\cite{NonlinearExcitonsTaghizadehPedersen,NonlinearBSERauwolfHolzer,NonlinearBSERuanLouie},
but a general analytical expression to arbitrary order in the response remains elusive at present.
An alternative approach for studying electronic excitations propagates the electronic many-body system in real time and Fourier transforms the electric dipole to obtain nonlinear optical responses. 
In a Green's function framework, the exact propagation can be described by the Kadanoff-Baym equations~\cite{KBTextbook,NEGFStefanucciVanLeeuwen,RTGWPerfettoStefanucci,KBESemkatBonitz,KBEJunctionsTuovinenStefanucci,CCvsGFReevesVlcek} (KBEs).
The self-energy in KBEs is time non-local, which increases the computational effort considerably.

In order to reduce the computational effort of KBEs, we apply the real-time Bethe-Salpeter equation approach\cite{RTBSEAttaccalite,RTBSEAttaccaliteDucastelle,Chan2021,RTBSEPerfettoStefanucci,RTBSEHouQiu,RTBSESangalli}
(RT-BSE), which uses time local approximations
for the self-energy
to arrive at algorithms with computational cost
comparable to TDDFT with hybrid functionals. Specifically, the screened
exchange\cite{GWHedin,COHSEXCasida,COHSEXBetzinger} (SEX) approximation is applied.

RT-BSE allows for evaluation
of non-linear effects, such as second harmonic generation or
optical rectification\cite{Boyd2008,Chan2021}. In the time propagation scheme, the
non-linear effects are present to arbitrary order. For these effects,
description of excitons is again crucial for accurate determination of the absorption/emission spectra. 

We report an implementation of the RT-BSE approach for molecular systems in the CP2K
\cite{CP2KReview2020}
 program. We
performed a benchmark on standard test set of organic molecules~\cite{CASPT2SchreiberThiel} comparing excitation energies from
the real time (RT) approach to the ones from the linear response\cite{RTBSEAttaccalite,ExcitationsReviewOnidaRubio,AbInitioSpectraRohlfingLouie,BSEBlaseLoos} (LR) approach.
Note that LR-BSE and RT-BSE approaches are equivalent in the linear response limit (see Ref.~\citenum{RTBSEAttaccalite} or Supporting Information~1
(SI~1)).
Going beyond  the linear response limit, we study non-linear
optical phenomena in molecules, specifically we
investigate the emergence of second harmonic generation\cite{Boyd2008} and
optical rectification in the
non-centrosymmetric cysteine molecule.

\section{Equation of Motion for the Density Operator within the Real-Time Bethe-Salpeter Approach}

To study the interaction of a many-electron system with light, we follow the RT-BSE approach
described in Ref.~\citenum{RTBSEAttaccalite}, which employs the von Neumann equation
\begin{align}
	\frac{\partial \hat \rho(t)}{\partial t} = \frac{-\ci}{\hbar}[\hat H^\text{eff}(t), \hat\rho(t)] 
	\label{eq:EoMDensity}
\end{align}
to compute the time evolution of the one-particle density operator  $\hat \rho (t) $. 
The time-dependent effective one-particle Hamiltonian is 
\begin{align}
    \hat{H}^\text{eff}(t) =&\; \hat h^{G_0W_0} + \hat U(t)  
    + \hat{V}^\text{H}[\hat{\rho}(t)] - \hat{V}^\text{H}[\hat{\rho}(0)] 
	+ \hat{\Sigma}^\text{SEX}[\hat\rho(t)] - \hat{\Sigma}^\text{SEX}[\hat\rho(0)]
    \label{eq:SEXEoM}
\end{align}
where  $\hat{h}^{G_0W_0}$ is the static effective one-particle Hamiltonian operator obtained from a preceding DFT plus $G_0W_0$ calculation\cite{Reining2018,Golze2019},
\begin{align}
	\hat h^{G_0W_0}  = \sum_n 
     \ket{\psi _ n} \epsilon^{G_0W_0}_n\bra{\psi_n}
\end{align}
where $\ket{\psi_n}$ is a Kohn-Sham DFT (KS-DFT) orbital and $\epsilon^{G_0W_0}_n$ is the corresponding $G_0W_0$ quasiparticle energy. 
Note that the RT-BSE scheme can also be started from $GW$ schemes including self-consistency, for example from $GW$ with eigenvalue-selfconsistency in $G$ (ev$GW_0$) which might align better with  quasiparticle energies from higher-level theories~\cite{Golze2019,Veril2018,Schambeck2024,Knysh2024}.

 $\hat{V}^\text{H}$
is the Hartree mean-field operator
and $\hat{\Sigma}^\text{SEX}$ is the SEX self-energy operator, which are both specified
in more detail when rewriting the equation of motion in an atomic Gaussian basis.
In contrast to Ref.~\citenum{RTBSEAttaccalite}, the Coulomb-hole
self-energy is not included, since it is exactly canceled in \eqref{eq:SEXEoM} for static screened potential fixed
at $G_0W_0$ or ev$GW_0$ screening.
The dynamics starts at $t=0$, and $\hat{\rho}(0)$ is the initial density operator, 
\begin{align}
 \hat{\rho}(0) =    \sum_n  \ket{\psi _ n} f_n\bra{\psi_n}\,,
\end{align}
where $f_n\in \{0,1\}$ is the occupation of the Kohn-Sham orbital~$n$ from the KS-DFT calculation. 

$\hat U(t)$ is the time-dependent external field operator. In order to excite the system, we implement two options. First option is to apply an external time
dependent potential $\hat U(t)$ in form of electric field in length gauge \cite{LengthGaugeMattiatLuber,LengthGaugeDitlerLuber}
\begin{align}
	\hat U(t) = e E(t) \bm \epsilon \cdot  \hat r
\end{align}
where $E(t)$ is the time-dependent electric-field amplitude,
$\bm \epsilon$ is the electric field polarization direction and $\hat r$
is the position operator.

The second option is the delta kick\cite{RTBSEAttaccalite,LengthGaugeMattiatLuber,TDDFTYabanaBertsch}, which is
described in more detail in the SI~2.
In essence, the delta kick transforms the equilibrium density matrix $\hat \rho(0)$ to a
non-equilibrium one via unitary transformation
\begin{align}
    \hat \rho(0) \to \hat \rho' = \eu^{(-\ci e/\hbar) I \bm \epsilon \cdot \hat r} \,\hat \rho(0)\,
    \eu^{(\ci e/\hbar) I \bm \epsilon \cdot \hat r}
    \label{eq:delta_kick}
\end{align}
where $I$ is the scale of the kick and $\bm \epsilon$ is the polarisation of the kick.
This transformation is associated with electric field with profile
$\bm E (t)= I \bm \epsilon \delta (t)$, where $\delta (t)$ is the Dirac delta function, hence
the name of the excitation scheme.

We compute the observables from the Fourier transform of the time-dependent dipole moment associated with the dynamics of the density matrix\cite{TDDFTYabanaBertsch,TDDFTMuellerSierka}
(see SI~3 for details). The Fourier transform
of the dipole moments is refined using the Padé approximant interpolation\cite{PadeFTBrunerLopata,PadeFTMattiatLuber,Kick2024,Leucke2025},
which allows the use of a fine frequency grid.

\section{Real-time Bethe-Salpeter in a Gaussian Basis}

We represent all operators appearing in the effective one-particle Hamiltonian \eqref{eq:SEXEoM} as matrices in the basis of atom-centered Gaussian orbitals $\ket{\phi _ \mu}$.
Since this is a nonorthonormal basis set, the overlap matrix $\bm S$ with elements $S_{\mu \nu} = \braket{\phi _ \mu}{\phi _ \nu}$
enters the equation of motion (see SI~4 for details) as follows
\begin{align}
	\frac{\partial \bm \rho}{\partial t} = \frac{- \mathrm i}{\hbar} \left (
		\bm S^{-1} \bm H^\text{eff} (t) \bm \rho (t) - \bm \rho (t) \bm H^\text{eff} (t) \bm S^{-1}
	\right )
	\label{eq:EoMBasis}
\end{align}
where $\bm H^\text{eff}$ is the matrix with elements of the respective operators introduced in Eq.~\eqref{eq:SEXEoM},
\begin{align}
    H^\text{eff}_{\mu\nu} (t)= \bra{\phi_\mu} \hat{H}^\text{eff}(t) \ket{\phi _ \nu}
    =
    h^{G_0W_0}_{\mu\nu} + U_{\mu\nu} (t) +V^\text{H}_{\mu\nu} [\bm\rho(t) -\bm\rho(0)] +
    \Sigma ^ \text{SEX}_{\mu\nu} [\bm \rho (t)-\bm \rho (0)]
\end{align}
and $\bm \rho(t)$ is the time-dependent density matrix with elements
$\rho _ {\mu\nu}(t)$ that satisfy
\begin{align}
	\hat \rho(t) = \sum _ {\mu,\nu} \rho _ {\mu\nu} (t)\ket{\phi _ \mu} \bra{\phi _ \nu} \: .
\end{align}
These definitions ensure that time-dependent observables $X(t)$ of an operator $\hat X$ have
expectation values represented by traces without the presence of the overlap matrix, i.e.
\begin{align}
X(t) = 	\trace \left ( \hat \rho(t) \hat X \right ) = \trace \left ( \bm \rho(t) \bm X \right ) \: ,
\end{align}
where $\bm X$ has matrix elements $X_{\mu\nu}=\bra{\phi_\mu} \hat{X} \ket{\phi _ \nu}$ as discussed in more detail in the SI~4.

An initial density at time $t$ is  propagated using the enforced time reversal symmetry (ETRS)
scheme\cite{PropagatorsCastroRubio}
\begin{align}
	\bm \rho (t + \Delta t) =
	\exp \left (- \frac{\mathrm i}{\hbar} \bm S^{-1} \bm H^\text{eff} (t + \Delta t) \frac{\Delta t}{2} \right )
	\bm \rho \left ( t + \frac{\Delta t}{2} \right )
	\exp \left (\frac{\mathrm i}{\hbar} \bm H^\text{eff} (t + \Delta t) \bm S^{-1} \frac{\Delta t}{2} \right ) \: ,
	\label{eq:ETRS}
\end{align}
where
\begin{align}
	\bm \rho \left (t + \frac{\Delta t}{2} \right ) =
	\exp \left (- \frac{\mathrm i}{\hbar} \bm S^{-1} \bm H^\text{eff} (t) \frac{\Delta t}{2} \right )
	\bm \rho ( t )
	\exp \left (\frac{\mathrm i}{\hbar} \bm H^\text{eff} (t) \bm S^{-1} \frac{\Delta t}{2} \right ) \: .
\end{align}
The matrix exponential in the equations above can be determined using the
Baker-Campbell-Hausdorff\cite{PropagatorsCastroRubio,PropagatorsORourke}
formula or by exact diagonalisation, more information is given in SI~5.

As can be seen from \eqref{eq:ETRS}, ETRS is an implicit scheme - the effective Hamiltonian at $t+\Delta t$
depends on $\bm \rho (t + \Delta t)$ that we  aim to compute. In our implementation, the implicit
dependence is solved by self-consistent (SC) iterations of the density matrix. Specifically, in iteration $n$,
we evaluate the effective Hamiltonian based on density matrix from previous iteration $\bm \rho _ {n-1}$, i.e.
\begin{align}
	\bm H^\text{eff} _n (t+\Delta t) = \bm H^\text{eff} _n [\bm \rho_{n-1}(t + \Delta t)](t+\Delta t)
\end{align}
with initial guess $\bm \rho_{0} (t + \Delta t)$ using $\bm H^\text{eff}[\bm \rho (t + \Delta t/2)](t + \Delta t)$
for the propagation from $\bm \rho (t + \Delta t/2)$. The SC criterion is checked as follows: For consecutive
density matrices in the SC iterations, element-wise difference is evaluated. The maximum absolute value of the
difference is noted. If this maximum is smaller than the SC threshold, the iterations are considered converged.
Importantly, ETRS scheme conserves the idempotency of the density matrix, see SI~6 for more details.

The numerically most expensive part of the ETRS loop is to recalculate the effective Hamiltonian,
specifically to compute the SEX self-energy and Hartree matrix. In Gaussian orbitals,
these are determined as (see SI~7 for details)
\begin{align}
	V ^ \text{H}_{\mu\nu}[\bm \rho (t)] = \sum _ {\lambda,\sigma} (\mu \nu | \sigma \lambda) \rho _{\lambda \sigma} (t) \: ,
	\label{eq:hartree_full}
\end{align}
where
\begin{align}
	( \mu \nu | \sigma\lambda ) = \int \mathrm d ^ 3 r \mathrm d ^ 3 r' \phi ^ * _ \mu ( \bm r) \phi _ \nu (\bm r)
	\frac{e^2}{4 \pi \epsilon _ 0|\bm r - \bm r'|} \phi ^ * _ \sigma (\bm r') \phi _ \lambda (\bm r') \: ,
\end{align}
and
\begin{align}
	\Sigma ^ \text{SEX}_{\mu\nu} [\bm \rho (t)] = - \sum _ {\lambda,\sigma} W_{\mu \lambda, \sigma \nu} \rho_{\lambda \sigma} (t) \: ,
	\label{eq:SEX_full}
\end{align}
where
\begin{align}
	W_{\mu \lambda, \sigma \nu} = \int \mathrm d ^ 3 r \mathrm d ^ 3 r' \phi ^ * _ \mu (\bm r) \phi _ \lambda (\bm r)
	W (\bm r, \bm r', \omega {=} 0) \phi ^ * _ {\sigma} (\bm r') \phi _ \nu (\bm r')
\end{align}
and $W(\bm r, \bm r', \omega{=}0)$ is the static screened Coulomb interaction, determined from $G_0W_0$. 
In both \eqref{eq:hartree_full} and \eqref{eq:SEX_full}, we employ the resolution of identity (RI)  to separate the 4-center kernels into 3-center ones\cite{Wilhelm2021,LSGWGramlWilhelm}.
Details are given in SI~8.

Note that the RT-BSE framework discussed in this work is closely related to real-time Hartree-Fock (RT-HF), which forms a fundamental component of real-time TDDFT when hybrid functionals are employed.
In RT-HF, the unscreened Coulomb interaction~$V$ is used in the exchange self-energy, in contrast to the screened Coulomb interaction~$W$ utilized in the screened-exchange self-energy~\eqref{eq:SEX_full}.
Importantly, $W$ only needs to be computed once at the beginning of the simulation. 
As a result, the computational cost of the time propagation is the same in RT-BSE, RT-HF, and real-time TDDFT with hybrid functionals.

\subsection{Observables}

The main observable recovered from the time propagation shown in this work is the time-dependent dipole moment
\begin{align}
	\bm \mu (t) = \Tr{\hat \rho (t) \hat \mu} = - e \Tr{\hat \rho (t) \hat r}
	\label{eq:moment_trace}
\end{align}

The absorption spectrum of the molecules is proportional to the isotropic part of the electric polarizability\cite{AbsorptionWu}
$\alpha(E)$
\begin{align}
	\alpha _ {jk}(\omega) = \frac{\mu _ j(\omega)}{E_k(\omega)}
	\label{eq:moment_response}
\end{align}
where $\mu_j(E)$ is the Fourier transform of the component $j$ of $\bm \mu(t)$ and $E_k(E)$ is the k-component
of the Fourier transform of electric field $\bm E(t)$. The isotropic part is then recovered by a trace
\begin{align}
	\alpha ^ \text{iso} (\omega) = \frac{1}{3} \Tr{\bm \alpha(\omega)} =
	\frac{1}{3} \sum _ {j\in \{x,y,z\}} \alpha _ {jj}(\omega) \label{alphaiso}
\end{align}

The polarizability determined by the RT-BSE method is compared with the LR-BSE polarizability, which is determined
from \cite{TDDFTUllrich}
\begin{align}
	\alpha ^ \text{iso}(\omega) = \frac{1}{3} \sum _ {n,j\in\{x,y,z\}} \frac{2 \Omega_n |\bra{\Psi _ n} \hat r _ j \ket{\Psi _ 0}|^2}{
		(\hbar\omega + i \eta) ^ 2 - \Omega _ n ^ 2} \: ,\label{alphaisoLRBSE}
\end{align}
where $n$ enumerates LR-BSE excitation energies $\Omega_n$, $\hat r_j$ is the position operator for position component $j$, $\eta$
is small artificial broadening, $\ket{\Psi _ 0}$ is the equilibrium state and $\ket{\Psi _ n}$ is the excitation state.

\section{Computational Details}

\subsection{Program Parameters}

The starting DFT calculation was carried out in the CP2K
suite, using the PBE0\cite{PBE0PerdewBurke, Adamo1999} functional with the
aug-cc-pVDZ basis set\cite{DunningBasis,DunningBasisKendallHarrison,BasisSetExchange2007,BasisSetExchange2019,BasisSetExchangeFeller}
as orbital basis set and aug-cc-pVTZ-RIFIT basis set\cite{DunningBasisRIWeigendHaettig} as auxiliary basis set
for the RI (see SI~9 for RI basis set convergence details).
The self-consistent iterations were converged so that the change in
density matrix elements between self-consistent field (SCF) iterations was less than $1.0 \times 10^{-7}$~at.u..
The energy cutoff for the plane-wave grid of the density matrix was 600~Ry.

The consecutive low-scaling $G_0W_0$ calculation\cite{Wilhelm2021} uses
a minimax grid\cite{MinimaxAziziGonze,Liu2016,Azizi2023} with
30~points in imaginary time and imaginary frequency and a truncation radius of 7~{\AA} in the RI
with the truncated Coulomb metric
(see SI~10 for effect of the cutoff radius).

The real-time propagation employs the ETRS self-consistent loop\cite{PropagatorsCastroRubio} with convergence
criterion on the maximum
absolute value difference of elements of two consecutive candidate density matrices in atomic
orbital basis being less than $10^{-7}$~atomic~units (at.u.).
For the presented results,
we calculate the matrix exponential using the BCH scheme with threshold $  10^{-14}$~at.u. for the resulting elements of the
transformed density matrix.

\subsection{Fourier Transform and Padé Approximant Refinement}

To make meaningful comparisons between the chosen molecules in both LR-BSE and RT-BSE, we
identified clear peaks present in both of the refined spectra obtained from the methods, fitted the peaks with Lorentzian
function\cite{TDDFTUllrich} and compared the  peak positions between the two methods.

The initial spectra are obtained by fast Fourier transform of the time trajectory of the dipole moment $\bm \mu(t)$
of each molecule. We propagate the dynamics for 20~fs, using 20~000 steps
of 1~as (timestep of 1~as is shown to be converged for cysteine molecule in SI~11).
Since the value of the dipole moment at the end is different from the initial value, we apply
artificial exponential damping $\gamma$\cite{TDDFTYabanaBertsch,TDDFTMuellerSierka} during post-processing,
\begin{align}
	\bm \mu(\omega) = \int _ 0 ^ \infty \mathrm{d}t\, \eu^{\ci (\omega + \ci \gamma) t} \bm \mu(t) \: ,
\end{align}
which acts as a window function of the Fourier transform and effectively broadens the spectrum. A value of $\gamma = 0.2\:\text{fs}^{-1}$ was chosen for the damping in order
to reduce the amplitude of dipole moment oscillations by $\eu^{-4}$ by the end
of the 20~fs propagation.
Further details are given in the SI~3.

We obtain  refined spectra  by fitting a Padé approximant to $\bm \mu(E)$ and interpolating
on a grid of energies from 0 to 20~eV, with 20~meV steps. The refinement uses $N/2$ Padé parameters,
where $N$ is the total number of the points considered for the fit,
consistently with previous work\cite{PadeFTBrunerLopata}. Points of energy up to 200~eV from the initial
spectra are used for the fitting.

\begin{figure}[h]
	\centering
	\includegraphics[width=\columnwidth]{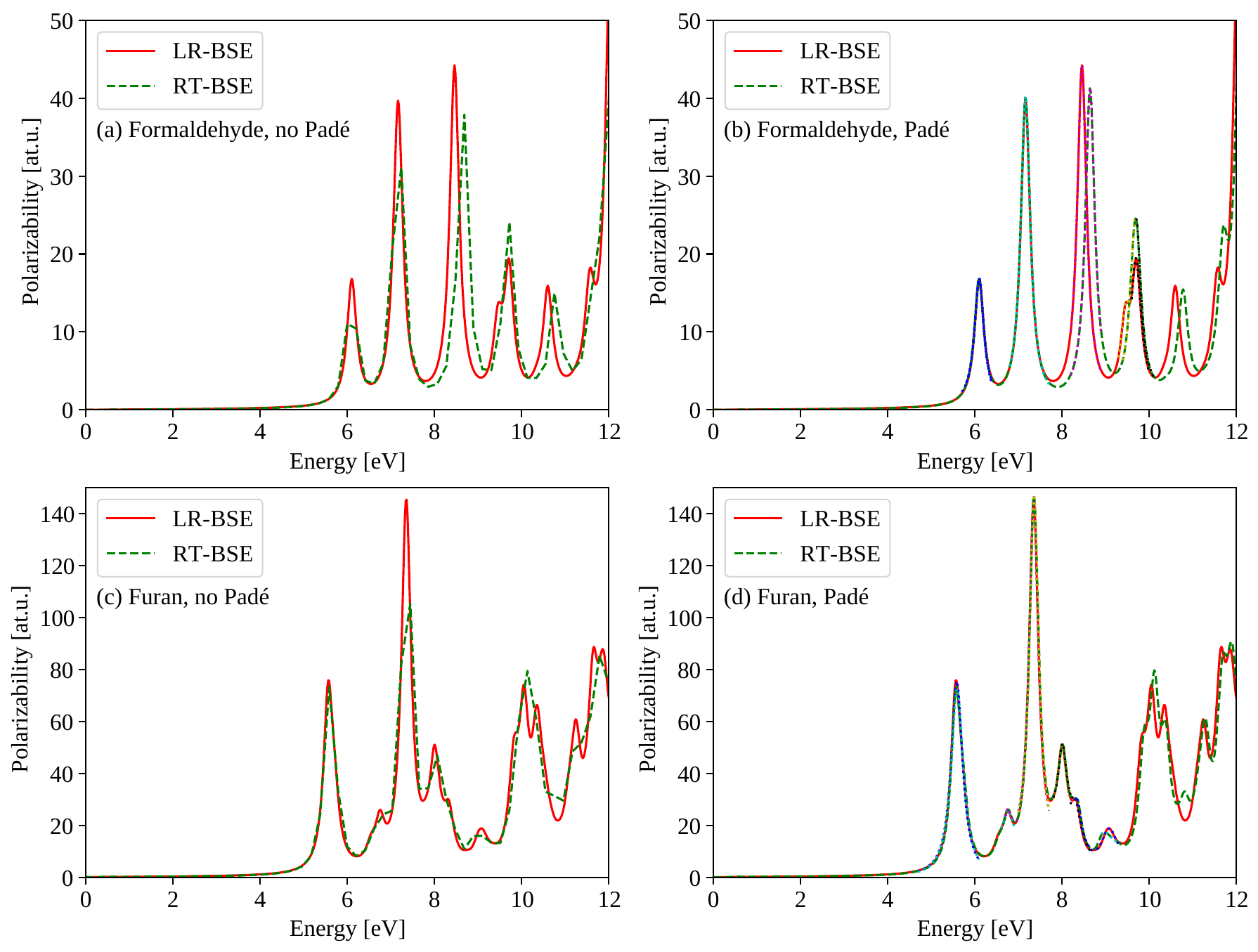}
	\caption{Isotropic polarizabilities of formaldehyde (top row, (a) and (b)) and
	furan (bottom row, (c) and (d)) computed from  RT-BSE propagation
	scheme~\eqref{alphaiso}, compared with the linear response (LR-BSE) approach~\eqref{alphaisoLRBSE}.
	The left column ((a) and (c)) shows spectra recovered
	from Fourier transform of the dipole moment time series without Padé interpolation,
	right column ((b) and (d)) shows
	spectra with Padé interpolation.
	The blue and yellow dotted lines in the (b) and (d) represent
	Lorentzian fits used to determine peak positions.}
    \label{fig:spectrum_example}
\end{figure}

\section{Accuracy with Respect to LR-BSE}

\figref{fig:spectrum_example}\,(a) and~(c) show RT-BSE isotropic polarisability spectra (green)
directly obtained from the Fourier transform of the dipole, Eq.~\eqref{alphaiso}.
Here, the energy resolution is only $\sim 0.2$~eV due to the total propagation time of 20~fs.
After applying Padé interpolation\cite{PadeFTBrunerLopata,PadeFTMattiatLuber,Kick2024,Leucke2025}, the energy resolution can be increased to arbitrary  precision, as shown in \figref{fig:spectrum_example}\,(b) and~(d).
Now, the RT-BSE and LR-BSE are indistinguishable by eye for energies up to 8 eV. 
For higher energies of formaldehyde in \figref{fig:spectrum_example}\,(b), peak positions between LR-BSE and RT-BSE differ by up to $\sim 0.2$~eV. 
We attribute the increasing deviation for peaks at higher energies to the analytic continuation used in low-scaling $GW$~\cite{Wilhelm2021} which serves as input for RT-BSE and which gets inaccurate for quasiparticle energies away from HOMO and LUMO (for LR-BSE, we use standard-scaling $GW$~\cite{GWWilhelmHutter} with 300 frequency points, where the accuracy of analytic continuation is guaranteed in a larger energy window below HOMO and above LUMO).

\begin{figure}[ht]
	\centering
	\includegraphics[height=0.49\columnwidth]{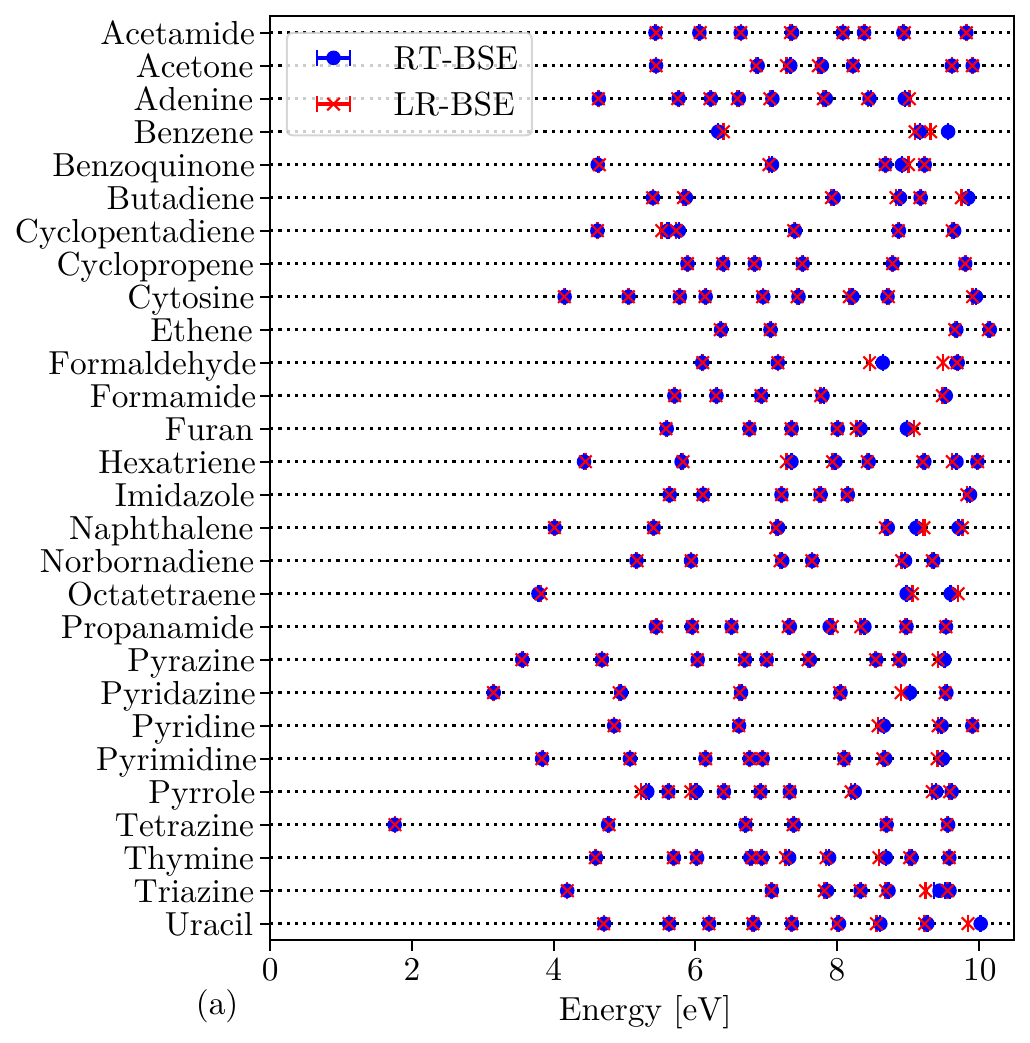}
	\includegraphics[height=0.49\columnwidth]{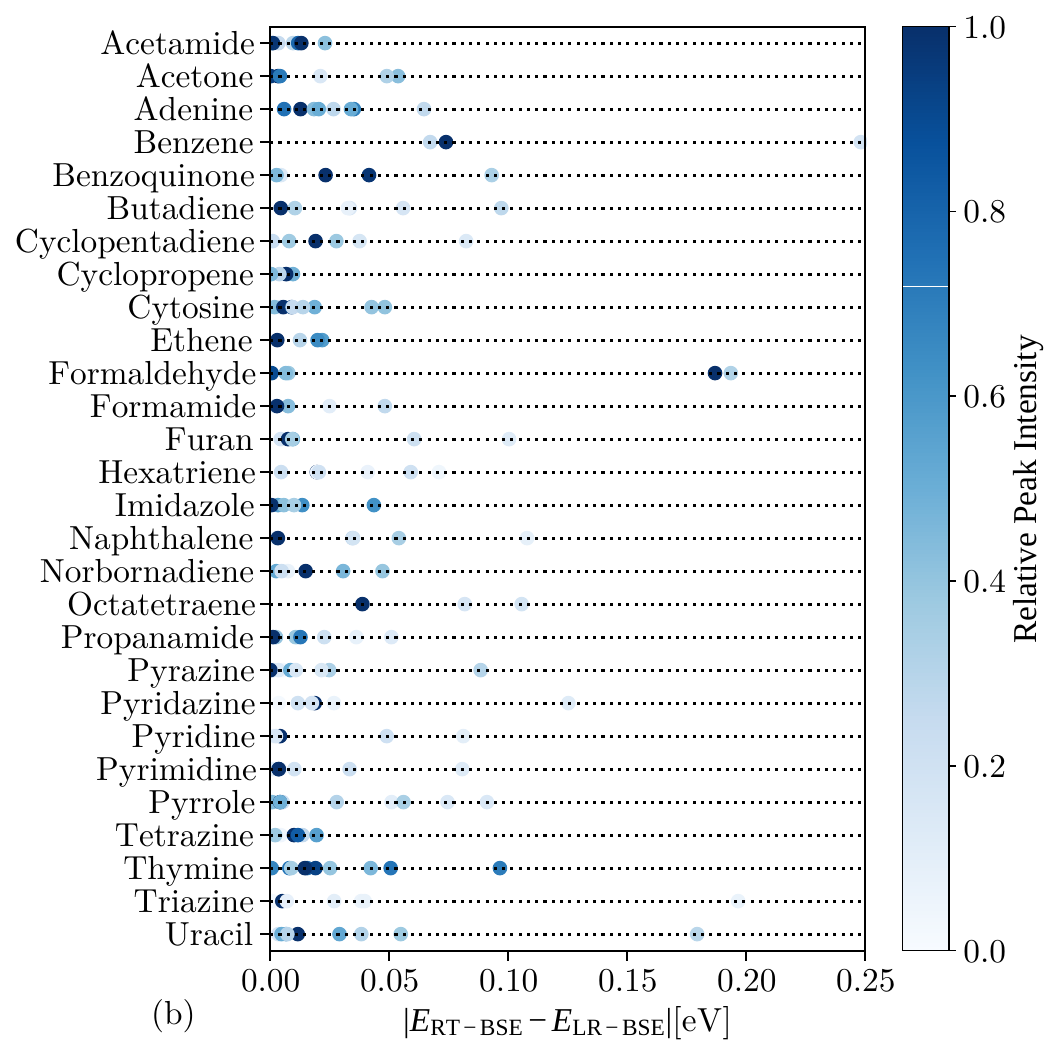}
	\caption{(a): Peak positions obtained from fits to LR-BSE and
	RT-BSE spectra. We observe excellent agreement in general.
	(b): Difference between RT-BSE and LR-BSE peak positions, displaying
	also the relative peak intensity as color --
	more pronounced peaks in the spectrum (below 10 eV) have more pronounced color. The peaks
	that differ the most between RT-BSE and LR-BSE ($>0.15$~eV difference)
	have usually low relative intensity, i.e.,  they are not the
	dominant feature of the molecular spectrum.
    }
    \label{fig:peak_positions}
\end{figure}

A systematic comparison of the peak positions computed from RT-BSE and LR-BSE across all
molecules in Thiel's set is presented in
\figref{fig:peak_positions}. 
In \figref{fig:peak_positions}a we observe that the peaks of RT-BSE and LR-BSE
are in excellent agreement; the average absolute deviation is 30~meV and the largest deviation is around 200~meV.
%
\figref{fig:peak_positions}a shows the peak position, independent of the peak height.
\figref{fig:peak_positions}b shows the absolute difference of peak positions between RT-BSE and LR-BSE, with the color bar indicating the peaks' relative intensity.
For peaks exceeding 50\% relative peak amplitude, the average absolute deviation is reduced to 20~meV.
Moreover, from \figref{fig:peak_positions}b, formaldehyde can be identified as the outlier
regarding numerical precision of RT-BSE, showing the largest deviation of
200~meV between RT-BSE and LR-BSE for peaks with $>$\,50\,\% relative peak intensity.
Nevertheless, as seen in Fig.~\ref{fig:spectrum_example}(b),
the RT-BSE and LR-BSE spectra are indistinguishable by eye up to 8~eV,
underscoring the numerical precision of our RT‑BSE implementation.

\section{Nonlinear-Optical Effects in Cysteine}
\label{sec:shg}
\begin{figure}[ht]
    \centering
    \includegraphics[width=\columnwidth]{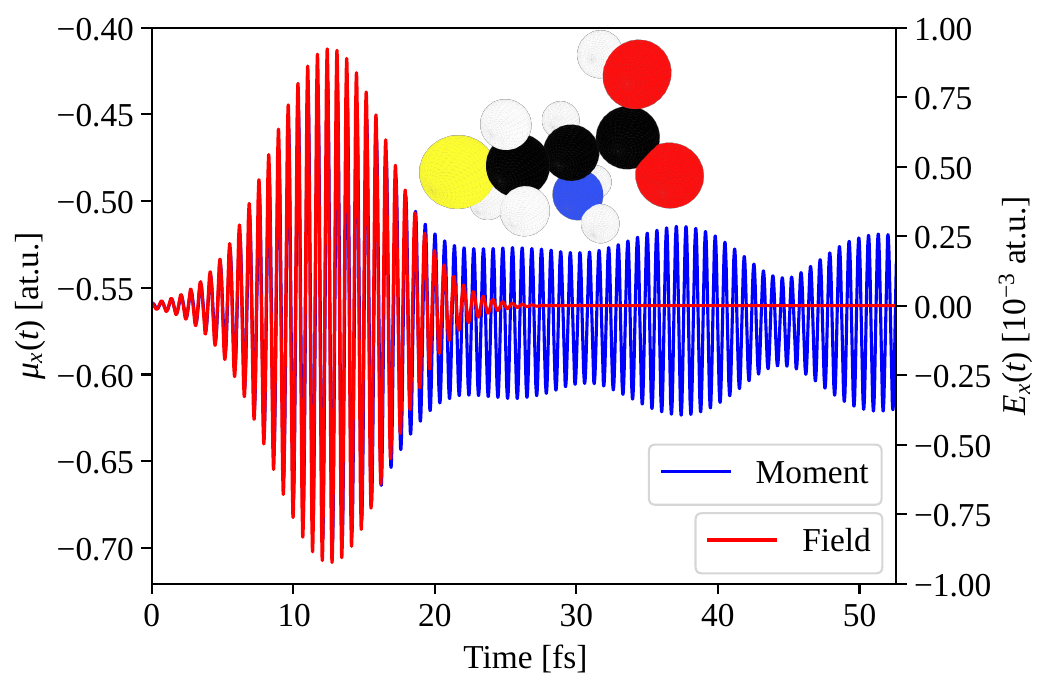}
    \caption{Real time pulse (in red) and induced dipole moment oscillations (in blue) of the
    cysteine molecule (molecular geometry shown), along the Cartesian $x$ direction.
    Oscillations of the amplitude after the initial pulse hint at presence of multiple frequencies.
    Furthermore, the center of the oscillations after the pulse is shifted,
    implying average static polarization of the molecule.}
    \label{fig:cysteine_trace}
\end{figure}

In linear optics, the response~$\bm\mu(t)$ of the molecule to light is proportional to the electric field amplitude~$E_0$ of the incoming light wave.
In nonlinear optics, the response~$\bm\mu(t)$ includes higher-order terms of~$E_0$ that become significant for large~$E_0$. 
Our chosen molecule to study nonlinear effects is cysteine, an amino acid, see molecular structure sketched in the inset of Fig.~\ref{fig:cysteine_trace}. 
Cysteine lacks inversion symmetry and therefore allows for second harmonic generation~\cite{Boyd2008}. 
Furthermore, being an amino acid, our choice of cysteine demonstrates the applicability of our method to
a molecule present in proteins.
The linear absorption spectrum of cysteine is shown in Fig.~\ref{fig:cysteine_trace} with an absorption peak at 6.0\,eV. 
We study nonlinear effects in the electric field amplitude using an electric-field pulse
\begin{align}
    E(t) = E_0 \cos(\omega_0 t) \,\eu ^ {-  {(t - t_0)^2}/({2 \sigma ^ 2})}
\end{align}
where $t _ 0 = 12.6$~fs, $\sigma=4.2$~fs, $\hbar \omega_0= 6.0$~eV and the field amplitude $E_0$ was varied between
different pulses. We use a simulation time of 52.6~fs,
in order to access the oscillations without the effect of the pulse field.
 For the Fourier transform from time to frequency, we use a Gaussian window function
(see SI~3.3 for details) centered at 32.6~fs with
spread of 4.0~fs.
The time trace of the field, $E(t)$ and  dipole moment~$\mu_x(t)$ computed from RT-BSE (using \eqref{eq:moment_trace}) are shown in \figref{fig:cysteine_trace}.

\begin{figure}[ht]
    \centering
    \includegraphics[height=0.3\columnwidth]{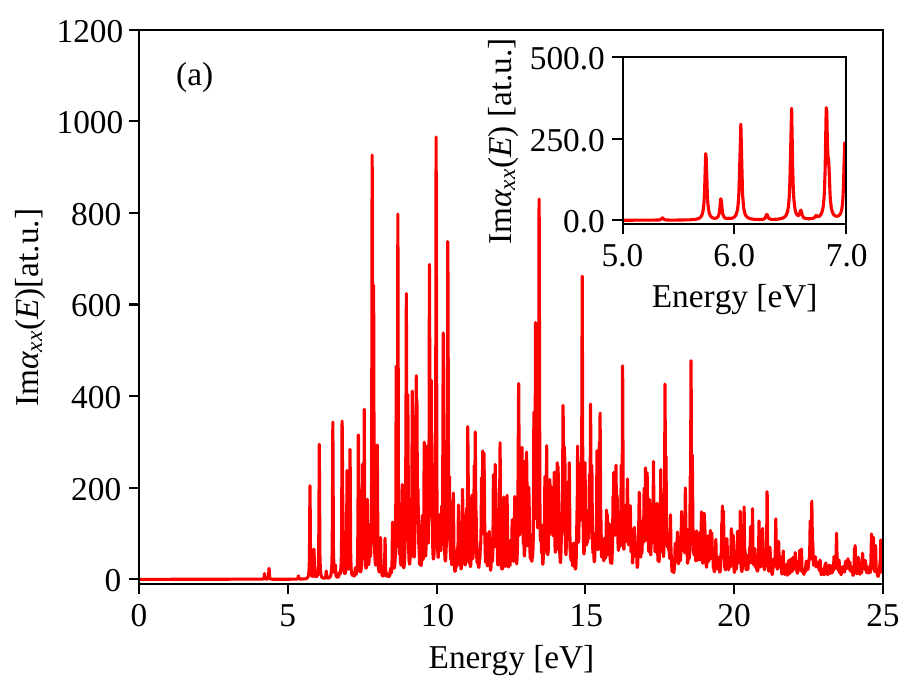}
    \includegraphics[height=0.3\columnwidth]{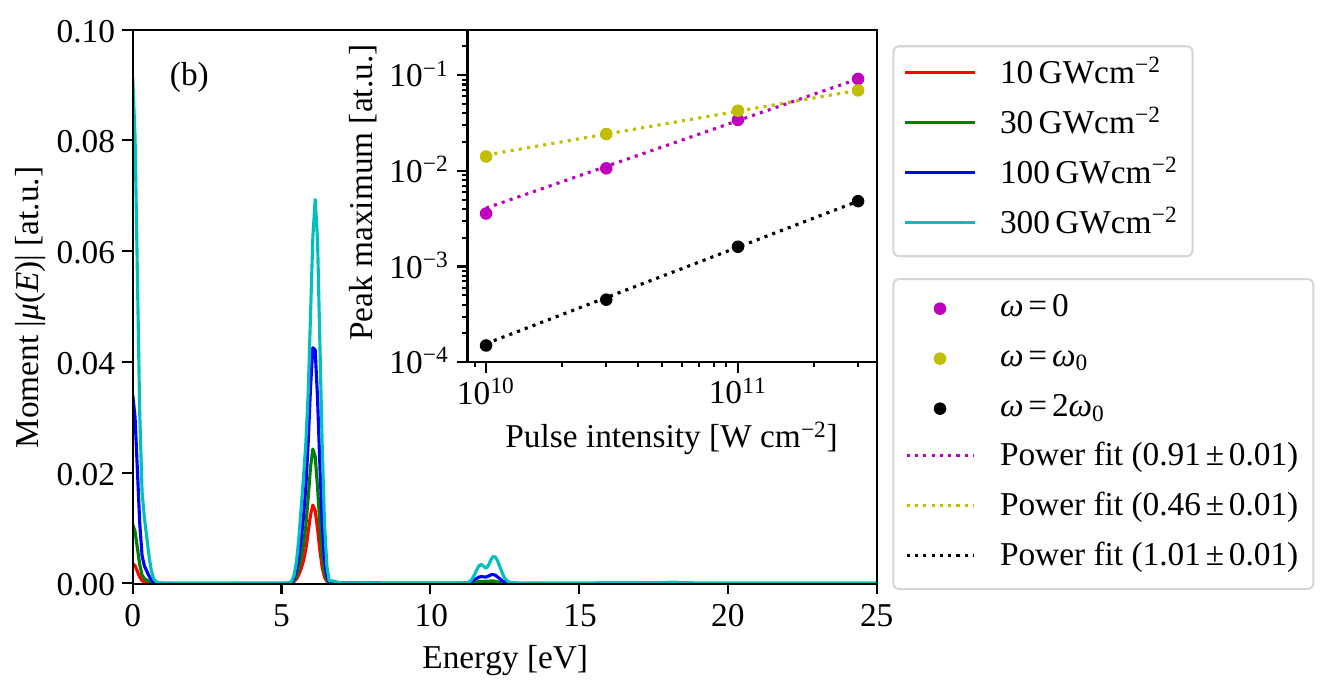}
    \caption{LR-BSE polarizability spectra (left) and magnitude of the Fourier transform
	of moment oscillations (right) for various
    intensities of the applied pulse. The pulse is centered around 6.00~eV, where the linear response
    spectrum has a low energy but reasonable intensity peak. The 2nd harmonic peak around 12~eV emerges with increasing
    intensity of the pulse.
    Note that there is also a static shift (optical rectification) of the dipole moment,
    increasing with the pulse intensity. In the inset,
    power fit to the maximum of each 0th, 1st and 2nd harmonic peak is displayed, showing an 
    approximately linear dependence on the intensity for the 0th and 2nd harmonic peaks and
    square root dependence for the 1st harmonic peak.}
    \label{fig:cysteine_ft_moment}
\end{figure}

Resulting energy dependent moment magnitudes are shown in \figref{fig:cysteine_ft_moment}.
We identify the linear response peak in \figref{fig:cysteine_ft_moment}a at $\sim$\,6\,eV and the 
second harmonic peak at $\sim$\,12\,eV. 
Our pulse has a non-zero spectra width $\sim 1$~eV, so several resonance energies~$\Omega_n$ get excited, see \eqref{eq:moment_response}.
Each linear-response peak gets broadened by the window function, so that only a single, broadened linear-response peak at $\sim$\,6\,eV is visible. 
For the second-order response at $\sim$\,12\,eV, we observe two distinct peaks.
This is because the spacing of the second-order response peaks is doubled compared to the linear response peaks. 
The maximum of the second harmonic peak scales
linearly with the intensity of the applied pulse $I = \epsilon_0 |E_0|^2$,
while the linear response peak maximum scales with $\sqrt I$.
Both of these observations are in agreement with the textbook
knowledge about linear and nonlinear optics~\cite{Boyd2008}.
We also identified an increasing peak at zero frequency. Since the maximum of this peak
shows the same scaling as the second harmonic peak (see inset of \figref{fig:cysteine_ft_moment}),
we attribute this peak to optical rectification phenomenon -- a second order effect\cite{Boyd2008}.
Summarizing, the study of nonlinear optical properties shown here shows the capabilities of RT-BSE applied to organic molecules. 

\section{Conclusions}

We have implemented the real-time Bethe-Salpeter equation approach via propagation of the density
matrix with screened-exchange self-energy in CP2K. We benchmarked the accuracy of the implementation with
respect to linear response Bethe-Salpeter equation approach, showing an  mean average absolute deviation  of
peak positions  in the isotropic polarizability of only 30~meV.
The application potential of the method was demonstrated by showing the emergence of the
second harmonic peak and optical rectification in the dipole moment magnitude of the amino acid cysteine.
The RT-BSE method is an important alternative approach to LR-BSE,
expanding it to capture non-linear optical effects
and allowing for explicit control of the pulse shape, both of which are inaccessible in
standard LR-BSE.

\section{Data and Code Availability}

Code described in this work is part of the free and open source CP2K\cite{CP2KReview2020} suite. The data described in the work
are freely available at NOMAD\cite{NOMADData} repository,
Github repository\cite{GithubData} and
Zenodo repository\cite{ZenodoData}.

\begin{acknowledgement}
The authors acknowledge frequent and insightful discussions with M.~Graml.
We acknowledge the DFG for funding via the Emmy Noether Programme (project 503985532), CRC 1277 (project number 314695032, subproject A03) and RTG 2905 (project number~502572516). 
%
The authors gratefully acknowledge the computing time provided to them on the high performance computer Noctua 2 at the NHR Center PC2. These are funded by the Federal Ministry of Education and Research and the state governments participating on the basis of the resolutions of the GWK for the national high-performance computing at universities (www.nhr-verein.de/unsere-partner).

\end{acknowledgement}

\begin{suppinfo}
	\begin{itemize}
		\item Additional theoretical and methodological explanations (PDF)
        \begin{enumerate}
            \item LR-BSE and RT-BSE equivalence
            \item Delta kick for homogeneous excitation
            \item Dipole polarizability from dipole moment time series
            \item Working in overlapping basis
            \item Exponentiation methods
            \item Conservation of idempotency
            \item Hartree and COHSEX terms in Gaussian overlapping basis
            \item RI approximation
	    \item RI auxiliary basis convergence
	    \item RI cutoff radius convergence
	    \item Timestep convergence in HHG of cysteine
	    \item Computational scaling
        \end{enumerate}
        \item Numerical results are available at NOMAD repository
		(\url{https://doi.org/10.17172/NOMAD/2025.04.17-1}),
		Github repository (\url{https://github.com/StepanMarek/RTBSEvsLRBSE}) and
		Zenodo repository (\url{https://doi.org/10.5281/zenodo.15235245})
	\end{itemize}
\end{suppinfo}

\bibliography{sources}

\end{document}


\input{supp_contents.tex}

\bibliography{sources}

%% file: supp_contents.tex
\section{LR-BSE and RT-BSE Equivalence}
\label{supp:lrbse_eq}

Before continuing to the specifics of the implementation, let us quickly show the equivalence of
the RT-BSE approach with the LR-BSE method in the linear regime.
In Ref.~\citenum{RTBSEAttaccalite}, authors show that the real-time
propagation of density matrix with COHSEX
self-energy in linear regime leads to equations are formally equivalent to standard Bethe-Salpeter equation.
Here, we briefly show a slightly less general way of deriving the Bethe-Salpeter equation in the
explicit form shown for example in Ref.~\citenum{BSELiuBlum},
again based on the COHSEX equation of motion~(1) in the main text.

Suppose that in equilibrium, the density matrix describing the effective single-particle states
is $\hat{\rho}^0$
\begin{align}
    \hat{\rho}^0 = \sum _ m f_m \ket{m}\bra{m} \: ,
    \label{eq:equil_density}
\end{align}
where $\ket{m}$ are the equilibrium Hamiltonian ($\hat{H}^0$) eigenstates
\begin{align}
    \hat{H}^0 \ket{m} = \hbar \omega _ m \ket{m} \:,
\end{align}
where $\hbar \omega_m$ is the energy eigenvalue, the eigenvectors are orthonormal
$\braket{m}{n} = \delta_{mn}$ and $f_m$ is a limit of the
Fermi distribution for zero temperature - if the highest occupied molecular orbital of the system
is $M$, then
\begin{align}
    f_m = \begin{cases}
        1 \: \mathrm{if} \: m \leq M \\
        0 \: \mathrm{if} \: m > M \\
    \end{cases} \: .
\end{align}

Now, suppose that the state of the system is changed so that the state is now described by density
matrix $\hat \rho(t) = \hat{\rho}^0 + \Delta \hat{\rho}(t)$.
Inserting into the equation of motion leads to
\begin{align}
    \frac{\partial \Delta \hat{\rho}}{\partial t} = \frac{\partial \hat{\rho}}{\partial t} =
    \frac{-i}{\hbar} \left [
	    \hat{H}^\text{eff}(t), \hat{\rho}^0 + \Delta \hat{\rho}(t)
    \right ] \: .
\end{align}

Without the external field $\hat{U}(t)$, the
only time-dependence in the Hamiltonian is present due to the changes in the density matrix, i.e.
\begin{align}
    \frac{\partial \Delta \hat{\rho}}{\partial t} &= 
    \frac{-i}{\hbar} \left ( \left [
        \hat{H}^0, \hat{\rho}^0
    \right ] + \left [
        \hat{H}^0, \Delta \hat{\rho}(t)
    \right ] + \left [
        \hat{V}^\text{Hartree}[\hat{\rho}(t)] - \hat{V}^\text{Hartree}[\hat{\rho}^0], 
        \hat{\rho}^0 + \Delta \hat{\rho}(t)
    \right ] + \right . \nonumber \\
    &+ \left . \left [
        \hat{\Sigma}^\text{COHSEX}[\hat{\rho}(t)] - \hat{\Sigma}^\text{COHSEX}[\hat{\rho}^0],
        \hat{\rho}^0 + \Delta \hat{\rho}(t)
    \right ]\right ) \: .
\end{align}

The first commutator is zero. For the resolution of the rest, it is useful to transform to the
basis of eigenstates of $\hat{H} ^ 0$
\begin{align}
    \frac{\partial \Delta \rho_{mn}}{\partial t} &= - \frac{i}{\hbar} \Biggl (
        \left [ \bm H ^ 0 , \Delta \bm \rho (t)\right ] _ {mn} + \nonumber \\
        &+ \sum_{p,q,r} (mp|rq) \Delta \rho_{qr}(t) (\bm \rho^0 + \Delta \bm \rho(t))_{pn} - 
        (\bm \rho^0 + \Delta \bm \rho(t))_{mp} (pn|rq) \Delta \rho_{qr}(t) - \nonumber \\
	&- \left . \sum_{p,q,r} W_{mq,rp} \Delta \rho_{qr}(t) (\bm \rho^0 + \Delta \bm \rho(t))_{pn} -
	(\bm \rho^0 + \Delta \bm \rho(t))_{mp} W_{pq,rn} \Delta \rho_{qr}(t)
    \right ) \: .
\end{align}

Now, we restrict ourselves to linear regime, i.e., we neglect higher than linear powers of
$\Delta \bm\rho (t)$. Furthermore, we substitute for $\bm \rho^0$ from \eqref{eq:equil_density}, which
leads to
\begin{align}
    \frac{\partial \Delta \rho_{mn}}{\partial t} &= - \frac{i}{\hbar} \Biggl (
        \hbar (\omega_m - \omega_n) \Delta \rho _ {mn}(t) + \nonumber \\
        &+ \sum _ {p,q} (mn|qp) \Delta \rho_{pq}(t) f_n -
        f_m (mn | qp) \Delta \rho_{pq}(t) - \nonumber \\
	&- \left . \sum_{p,q} W_{mp,qn} \Delta \rho_{pq}(t) f_n -
	f_m W_{mp,qn} \Delta \rho_{pq}(t)
    \right ) \: .
\end{align}

After some reordering, the equation of motion in linear response regime
for the perturbation $\Delta \bm \rho(t)$ is
\begin{align}
    \frac{\partial \Delta \rho_{mn}}{\partial t} = i(\omega_n - \omega_m) \Delta \rho_{mn}(t)
	- \frac{i}{\hbar} (f_n - f_m) \sum_{p,q} \left ( (mn|qp) - W_{mp,qn} \right ) \Delta \rho_{pq}(t)
    \label{eq:linres_eom_time}
\end{align}
By applying the Fourier transform, we obtain
\begin{align}
    \hbar \omega \Delta \rho_{mn}(\omega) = - \hbar (\omega_n - \omega_m) \Delta \rho_{mn}(E) + (f_n - f_m) \sum_{p,q} \left (
	(mn|qp) - W_{mp,qn} \right ) \Delta \rho_{pq}(\omega)
    \label{eq:linres_eom_energy}
\end{align}

This is as far as we can get for a general form of $\Delta \rho_{mn}$. We now choose a specific ansatz
for the structure of $\Delta \rho _ {mn}$ - we allow for it to only contain off-diagonal terms in
the occupied+unoccupied blocks. Specifically, we write
\begin{align}
	\Delta\hat{\rho} = \sum _ {mn} \left ( X_{nm} f_n (1-f_m) + Y_{mn} f_m(1-f_n) \right ) \ket{m}\bra{n}
    \label{eq:linres_off_diagonals}
\end{align}

\subsection{Linear Response Density via Delta-Kick}

In order to justify the ansatz \eqref{eq:linres_off_diagonals}, consider that the initial
non-equilibrium part of the density can result from applying the delta-kick,
described later on. The effect of the delta kick can be described as
\begin{align}
	\hat{\rho}(t=0^+) = \mathrm{e}^{-i \hat A} \hat \rho^0 \mathrm{e}^{i \hat A} =
	\hat \rho^0 + \Delta \rho(t=0)\: ,
\end{align}
where $\hat A$ is a Hermitian operator. Evaluating $\hat{\rho}(t=0^+)$ in the basis of
original Hamiltonian eigenstates $\ket{m}$ leads to
\begin{align}
	\bra{m} \hat{\rho}(t=0^+) \ket{n} = \sum _ {p,q} \bra{m} \mathrm{e} ^ {-i \hat A} \ket{p} \bra{p}
    \hat \rho^0 \ket{q} \bra{q} \mathrm{e}^{i\hat A} \ket{n}
\end{align}
Approximating the exponential and resulting matrix products to linear order in $\hat A$ leads to
\begin{align}
    \bra{m} \hat{\rho}(t=0^+)\ket{n} = \bra{m} \hat \rho^0 \ket{n} +
    \sum_{p,q} (-i) \bra{m} \hat A \ket{p} \bra{p} \hat{\rho}^0 \ket{q} \delta_{qn} +
    i \delta_{mp} \bra{p} \hat{\rho}^0 \ket{q} \bra{q} \hat A \ket{n}
\end{align}
Substituting from \eqref{eq:equil_density}
\begin{align}
    \bra{m} \hat{\rho}(t=0^+) \ket{n} = f_m \delta_{mn} +
    \sum _ {p} (-i) A_{mp} f_p \delta _ {pn} +
    i f_m \delta _ {mq} A_{qn} = f_m \delta_{mn} + i (f_m {-} f_n) A_{mn}
\end{align}
First of all, we note that the perturbed density matrix remains Hermitian
\begin{align}
	\bra{n} \hat{\rho}(t=0^+) \ket{m} ^ * &= (f_n \delta_{nm} + i (f_n - f_m) A_{nm})^* = \nonumber \\
	&= f_ m \delta _{mn} + i (f_m - f_n) A_{mn} = \bra{m} \hat{\rho}(t = 0^+) \ket{n}
\end{align}
since $\hat A$ is Hermitian.
The $f_m \delta_{mn}$ term corresponds to equilibrium density element $\bra{m} \hat{\rho}^0 \ket{n}$. 
The second term (corresponding to $\Delta{\rho}(0)$) is non-zero only if one of $m,n$ belongs to the
occupied subspace while the other belongs to the unoccupied subspace. We can change the second term
further
\begin{align}
    \bra{m} \Delta \hat{\rho} (0) \ket{n} &= i A _ {mn} (f_m - f_n) = 
    i A _ {mn} (f _ m - f _ m f _ n + f _ m f _ n - f _ n) = \nonumber \\
    &= i A _ {mn} f_m (1 - f _ n) - i A _ {mn} f _ n (1 - f _ m)
\end{align}

We can see that this form of density perturbation follows the ansatz \eqref{eq:linres_off_diagonals}
(with $Y_{mn} = i A _ {mn}$ and $X_{mn} = - i A_{nm}$).
To summarize, the off-diagonal occupied/unoccupied block structure of the density matrix perturbation
is a general feature of weak (linear response regime) unitary transformation of the density matrix 
away from the equilibrium density matrix.

\subsection{Casida Formalism}

Substituting to \eqref{eq:linres_eom_energy} from \eqref{eq:linres_off_diagonals}, we get
\begin{align}
    &\hbar \omega (X_{nm} (\omega) f_n(1-f_m) + Y_{mn}(\omega) f_m(1-f_n)) = \nonumber \\
    &- \hbar (\omega_n - \omega_m) (X_{nm}(\omega) f_n(1-f_m) + Y_{mn}(\omega) f_m(1-f_n)) + \nonumber \\
	&+ (f_n - f_m) \sum_{p,q} ((mn|qp) - W_{mp,qn}) (X_{qp} (\omega) f_q(1-f_p) + Y_{pq} (\omega) f_p(1-f_q))
\end{align}
We now notice
that only the values of $m,n$ for occupied + unoccupied orbital pairs lead to non-trivial equations.
Specifically, setting $i,j$ to stand for indices of occupied orbitals and $a,b$ for indices of
unoccupied orbitals, we have a set of two coupled equations
\begin{align}
    \hbar \omega X_{ia}(\omega) = \hbar (\omega_a - \omega_i) X_{ia}(\omega) + \sum_{j \: \text{occ.}, b \: \text{unocc.}}
	((ai|jb) - W_{ab,ji}) X_{jb} + ((ai|bj) - W_{aj,bi}) Y_{jb} \\
    \hbar \omega Y_{ia}(E) = -\hbar(\omega_a - \omega_i) Y_{ia}(E) - \sum_{j \: \text{occ.}, b \: \text{unocc.}}
	((ia|bj) - W_{ij,ba}) Y_{jb} + ((ia|jb) - W_{ib,ja}) X_{jb}
\end{align}

This is equivalent to the Bethe-Salpeter approach via the Casida equation. Specifically,
defining
\begin{align}
	A_{ia,jb} = \hbar (\omega_a-\omega_i)\delta_{ij} \delta_{ab} + (ia|jb) - W_{ij,ab} \nonumber \\
	B_{ia,jb} = (ia|bj) - W_{ib,aj} \: ,
\end{align}
one can write
\begin{align}
    \begin{pmatrix}
        \bm A & \bm B \\
        - \bm B^\dagger & -\bm A^\dagger \\
    \end{pmatrix} \begin{pmatrix}
        \bm X \\ \bm Y \\
    \end{pmatrix} = \hbar \omega \begin{pmatrix}
        \bm X \\ \bm Y \\
    \end{pmatrix}
\end{align}
which is consistent for example with notation in \citeauthor{BSELiuBlum}\cite{BSELiuBlum}

\section{Delta Kick for Homogeneous Excitation}
\label{supp:delta_kick}

Instead of a time resolved field pulse $\bm E (t)$, we might choose to apply a delta-like pulse
\begin{align}
    \bm E(t) = I \bm \epsilon \delta(t)
\end{align}
where $I$ is the scale of the delta pulse and $\bm \epsilon$ is the polarization\cite{LengthGaugeMattiatLuber}.
We resolve the effect of the delta pulse
on the density of states via explicitly integrating the equation of motion. Let $\hat \rho^-$ be the 
density matrix operator before the application of the pulse and $\hat \rho^+$ the density matrix operator
immediately after the application, i.e.
\begin{align}
    \hat \rho^\pm = \lim _ {\Delta t \to 0^+} \hat \rho(\pm \frac{1}{2} \Delta t)
\end{align}

Before the application of the pulse, $\hat \rho^-$ corresponds to the starting $\hat \rho_0$ provided by the
initial guess from DFT+$G_0W_0$. The equation of motion to zeroth order in $I$ is solved as
\begin{align}
    \hat \rho ^ + = \hat \rho ^ - - \frac{i}{\hbar} \int _ {-\Delta t/2} ^ {\Delta t/2} dt
    [\hat H_0, \hat \rho(t)] \approx \hat \rho ^ - - \frac{i}{\hbar} \Delta t [\hat H_ 0, \hat \rho ^ -]
\end{align}
where $\hat H _ 0$ is the time-independent DFT+$G_0W_0$ Hamiltonian. To the first order in $I$
\begin{align}
    \hat \rho ^ + = \hat \rho ^ - - \frac{i}{\hbar} \int _ {-\Delta t / 2} ^ {\Delta t / 2} dt
    [\hat H _ 0 + I \delta (t) \bm \epsilon \cdot \hat r, \hat \rho (t)] \approx \\
    \approx \hat \rho ^ - - \frac{i}{\hbar} \int _ {-\Delta t/2} ^ {\Delta t/2} dt
    \left [ \hat H _ 0 + I \delta (t) \bm \epsilon \cdot \hat r, \hat \rho ^ - - \frac{i}{\hbar}
    \int _ {-\Delta t/2} ^ {t} dt' [\hat H_0 + I \delta (t') \bm \epsilon \cdot \hat r, \hat \rho ^ -]\right ] = {}\\
    {} = \hat \rho ^ - - \frac{i}{\hbar} [\hat H _ 0, \hat \rho ^ -] \Delta t +
    \left ( \frac{-i}{\hbar} \right ) ^ 2 [\hat H _ 0, [\hat H _ 0, \hat \rho ^ -]]
    \underbrace{\int _ {-\Delta t/2} ^ {\Delta t/2} dt \int _ {-\Delta t / 2} ^ t dt'}_{(\Delta t)^2/2} + \\
    + \left ( \frac{-i}{\hbar} \right ) ^ 2 [\hat H _ 0, [I \bm \epsilon \cdot \hat r, \hat \rho ^ -]] \Delta t
    - \frac{i}{\hbar} [I \bm \epsilon \cdot \hat r, \hat \rho ^ -] + \\
    + \left ( \frac{-i}{\hbar} \right ) ^ 2 [I \bm \epsilon \cdot \hat r, [\hat H _ 0, \hat \rho ^ -]]
    \underbrace{\int _  {- \Delta t/2} ^ {\Delta t/2} dt \delta (t) \int _ {- \Delta t/2} ^ {t} dt' 1}_{\Delta t/2}
\end{align}

Now, we can notice that only the terms $\rho^-$ and $(-i)/\hbar[I \bm \epsilon \cdot \hat r, \hat \rho ^ -]$
are independent of $\Delta t$ and will therefore be finite even for $\Delta t \to 0$.

Similarly, we  derive that only term independent of $\Delta t$ to the second order in $I$ will be
\begin{align}
    \left ( \frac{-i}{\hbar}\right ) ^ 2
    [I \bm \epsilon \cdot \hat r, [I \bm \epsilon \cdot \hat r, \hat \rho ^ -]]
    \int _ {-\Delta t/2} ^ {\Delta t/2} dt \delta (t) \int _ {-\Delta t/2} ^ {t} dt' \delta(t') = {} \\
    {} = \left ( \frac{-i}{\hbar}\right ) ^ 2
    [I \bm \epsilon \cdot \hat r, [I \bm \epsilon \cdot \hat r, \hat \rho ^ -]]
    \int _ {-\Delta t/2} ^ {\Delta t/2} dt \delta (t) \theta(t) = \frac{1}{2} \left ( \frac{-i}{\hbar}\right ) ^ 2
    [I \bm \epsilon \cdot \hat r, [I \bm \epsilon \cdot \hat r, \hat \rho ^ -]]
\end{align}
and equivalently for higher orders, leading to expression replicating the
BCH theorem\cite{PropagatorsCastroRubio}
\begin{align}
    \hat \rho ^ + = \sum _ {n = 0} ^ \infty \frac{1}{n!} \left [
        \left ( \frac{-i}{\hbar} \right ) I \bm \epsilon \cdot \hat r, \hat \rho ^ - \right ] _ n =
    e ^ {- \frac{i}{\hbar} I \bm \epsilon \cdot \hat r} \hat \rho ^ -
    e ^ {\frac{i}{\hbar} I \bm \epsilon \cdot \hat r}
\end{align}

The expression in atomic basis follows exactly the BCH/exact diagonalisation scheme discussed later,
with substitution of $\bm H(t)$ by $I \bm \epsilon \cdot \hat r$.

\section{(Dipole) Polarizability from Dipole Moment Time Series}
\label{supp:observables_series}

The dipole polarizability $\alpha _ {mn} (\omega)$ ($m,n \in \{x,y,z\}$ are Cartesian directions)
is recovered from the time series of the dipole operator $\hat \mu _ m$ expectation value
and the time series of the applied field $E_n(t)$ as
\begin{align}
    \alpha_{mn} (\omega) = \frac{\mu_m(\omega)}{E_n(\omega)} = \frac{
    \int _ {t_0} ^ \infty dt \mathrm{e}^{\mathrm{i} \omega (t-t_0)} \mathrm{e}^{-\gamma (t-t_0)}
    \left \langle \hat \mu _ m(t) - \hat \mu _m (t_0)\right \rangle }{
    \int _ {t_0} ^ \infty dt \mathrm{e} ^ {\mathrm{i} \omega (t - t_0)} E_n(t)
    }
    \label{eq:moment_ft}
\end{align}
where $t_0$ is a reference time before which the density matrix is stationary (before application of
the excitation), $\hat \mu_m (t)$ is the dipole moment operator for Cartesian direction
$m \in \{x,y,z\}$, $\gamma$ is an artificial damping which
stabilizes the numerical Fourier transform\cite{TDDFTMuellerSierka}
and $E_n(t)$ is the external field along direction $n$.

\subsection{Optical Spectra from Real Time Moment Traces}

The absorption is given (quasi-classically) by the acceleration of the dipole moment, which introduces an extra
factor of $\omega$. The absorption of the field is associated with decreasing amplitude of the
resulting field, i.e. with the imaginary value of $\alpha$. Specifically, \cite{TDDFTMuellerSierka,LengthGaugeMattiatLuber,TDDFTCistaro,AbsorptionWu,TDDFTYabanaBertsch,ElectronicStructureMartin}
\begin{align}
	S(\omega) \propto \frac{1}{3} \omega \text{Tr} \left ( \text{Im} \left ( \bm \alpha (\omega) \right )\right ) \: .
\end{align}

\subsection{Spectrum Post-Processing}

The discrete Fourier transform is used to get the transform of the dipole moment oscillations. In
order to reduce numerical noise, several post-processing steps are applied.

Firstly, a damping envelope is applied to the real time signal.
For majority of the results, the
chosen envelope is the exponential damping, as shown in \eqref{eq:moment_ft},
which leads to effective broadening of the peaks of the oscillation, as can be seen from FT of
\begin{align}
    \int dt e ^ {i \omega t} \theta(t) \sin (\omega _ 0 t) e ^ {- \gamma t} = 
    \int _ 0 ^ \infty e ^ {i (\omega + i \gamma ) t} \sin(\omega _ 0 t) = \\
    = \frac{1}{2i} \int _ 0 ^ \infty dt \left [
        e ^ {i (\omega + \omega _ 0 + i \gamma ) t} -
        e ^ {i (\omega - \omega _ 0 + i \gamma) t}
    \right ] = \frac{1}{2} \left [ \frac{1}{\omega + \omega _ 0 + i \gamma} - \frac{1}{\omega - \omega _ 0 + i \gamma} \right ]
\end{align}
The sine was chosen as an oscillatory function that has 0 value at $t=0$. Besides numerical stabilization of the
peak positions by broadening, the exponential damping also leads to coinciding initial and final value of the moments, which removes some abnormal frequencies otherwise present in the FT.

The imaginary part of the FT is than the imaginary part of the expression in the brackets above,
which results in a Lorentzian peak
\begin{align}
    \frac{1}{2} \left [ \frac{-\gamma}{(\omega + \omega _ 0) ^ 2 + \gamma ^ 2} +
    \frac{\gamma}{(\omega - \omega _ 0) ^ 2 + \gamma ^ 2}\right ]
\end{align}

The polarizability depends also on the transform of the applied field. Therefore,
we set $t=0$ at the symmetry center of non-oscillatory envelope of the field - this leads to a purely
real Fourier transform. Numerically, this is achieved by wrapping the sample around certain
index of the signal data array.

\subsection{Other Possible Window Functions}
\label{supp:window_functions}

Note that one can also choose symmetric window functions such as Gaussian and symmetric exponential (Poisson) window function. For a general signal $f(t)$ with Fourier transform
\begin{align}
    f(\omega) = \sum _ i c_i \delta(\omega - \omega_i) \: ,
\end{align}
the effect of the window function $h(t)$ is understood through the convolution in Fourier space.
Specifically, 
\begin{align}
    \int _ {-\infty} ^ \infty f(t) h(t) e^{i \omega t}dt &= 
    \int _ {-\infty} ^ \infty dt \frac{1}{4 \pi ^ 2} \int _ {-\infty} ^ \infty d\omega
    \int _ {-\infty} ^ \infty d\omega' e^{-i\omega' t} e^{-i \omega'' t} e^{i \omega t}
    f(\omega') h (\omega'') = \\
    &= \int _ {-\infty} ^ \infty d\omega'' \int _ {-\infty} ^ \infty d\omega' f(\omega') h(\omega'')
    \frac{1}{2 \pi} \delta(\omega' + \omega'' - \omega) = \\
    &= \frac{1}{2 \pi} \int d\omega' f(\omega') h(\omega - \omega') = \\
    &= \frac{1}{2 \pi} \sum _ i c_i h(\omega - \omega _ i) \: ,
\end{align}
i.e. the Fourier transform of the window function replaces the delta peaks in the
spectrum $f(\omega)$ of original signal $f(t)$. For even more specific case of sine signal
\begin{align}
    f(t) = \sum _ i a_ i \sin(\omega _ i t)
\end{align}
we have
\begin{align}
    f(\omega) = i \pi \sum _ i a _ i (\delta (\omega - \omega _ i) - \delta(\omega + \omega _i))    
\end{align}
For exponential window $h(t) = e^{-\gamma|t|}$, so
\begin{align}
    h(\omega) = \frac{2 \gamma}{\omega ^ 2 + \gamma ^ 2}
\end{align}
and hence
\begin{align}
    \int _ {-\infty} ^ \infty f(t) h(t) e ^ {i \omega t} dt =
    i \gamma \sum _ i \frac{a_i}{(\omega - \omega _ i) ^ 2 + \gamma ^ 2} -
    \frac{a_i}{(\omega + \omega _ i) ^ 2 + \gamma ^ 2}
\end{align}

For Gaussian window function $h(t) = e^{-t^2/(2 \sigma^2)}$, $h(\omega) = \sigma \sqrt{2 \pi} e^{-\sigma^2 \omega ^ 2/2}$, so the sine signal produces
\begin{align}
    \int _ {-\infty} ^ \infty f(t) h(t) e ^ {i \omega t} dt =
    i \sigma \sqrt{\frac{\pi}{2}} \sum _ i a _ i e ^ {- \sigma^2 (\omega - \omega_i) ^ 2/2} -
    a _ i e ^ {- \sigma ^ 2 (\omega + \omega _ i) ^ 2 / 2}
\end{align}

In summary, both the exponential and the Gaussian window function produce a positive semi-definite imaginary spectrum in the positive frequency range for sine signal.

\section{Working in Overlapping Basis - Covariant and Contravariant Operator Representations}
\label{supp:overlapping_basis}

In an overlapping basis of state vectors, we have a set of vectors $\ket{\phi _ \mu}$
(we shall assume finite) which satisfy
\begin{align}
	\braket{\phi_\mu}{\phi_\nu} = S _ {\mu \nu}
\end{align}
where $S_{\mu \nu}$ is the so-called overlap matrix. Assume now that a new set of vectors
$\ket{\psi _ m}$ is constructed as linear combinations of $\ket{\phi_\mu}$,
\begin{align}
	\ket{\psi _ m} = \sum _ \mu C _ {\mu m} \ket{\phi _ \mu} , \braket{\psi _ m}{\psi _ n} = \delta _ {mn} \:,
\end{align}
where $\delta _ {mn}$ is the Kronecker delta. The condition on coefficients $C_{\mu m}$ can be then formulated
as matrix equation
\begin{align}
	\bm C ^ \dagger \bm S \bm C = \bm I \:,
\end{align}
where $\bm I$ is the identity matrix, or alternatively
\begin{align}
	\bm S ^ {-1} = \bm C \bm C ^ \dagger \:. \label{cc_sinv}
\end{align}

For representation of quantum mechanical operators as matrices, one usually employs the completeness relation
to resolve the matrix elements. In orthogonal basis, this takes form $\sum _ m \ket{\psi _ m} \bra{\psi _ m} = \hat{I}$,
where $\hat{I}$ is the identity operator. In the overlapping basis, this can be expressed as
\begin{align}
	\sum _ m \left ( \sum _ {\mu} C _ {\mu m} \ket{\phi _ \mu} \right )
	\left ( \sum _ {\nu} C _ {\nu m} ^ * \bra{\phi _ \nu}\right ) = 
	\sum _ {\mu, \nu} \ket{\phi _ \mu} \left ( \sum _ m C _ {\mu m} C _ {\nu m} ^ *\right ) \bra{\phi _ \nu} =
	\sum _ {\mu, \nu} \ket{\phi _ \mu} S ^ {-1} _ {\mu \nu} \bra{\phi _ \nu} \label{resId_overlap}
\end{align}

We crucially note that the last expression is independent of the choice of the orthonormal basis
and therefore represents the identity operator and associated completeness relation in the overlapping basis.

The presence of the overlap matrix in the completeness relation leads to difference in the representation of
operators, which we call covariant and contravariant representation. In the covariant representation, we represent
the operators by matrices
\begin{align}
	A _ {\mu \nu} = \bra{\phi _ \mu} \hat A \ket{\phi _ \nu} \:,
\end{align}
while in the contravariant representation, we use
\begin{align}
	\hat A = \sum _ {\mu, \nu} \ket{\phi _ \mu} A _ {\mu \nu} \bra{\phi _ \nu} \:.
\end{align}
The two representations are related by transformation
\begin{align}
	\bm A _ {\text{cov}} = \bm S \bm A _ \text{contra} \bm S \:.
\end{align}

\subsection{Reasoning Behind Representation Naming}

When the basis is transformed by a unitary transformation $\hat{U}$, such that
\begin{align}
	\ket{\phi' _ \mu} = \hat U \ket{\phi _ \mu}
\end{align}
the covariant representation of the operator transforms as
\begin{align}
	(\bm A ')_{\mu \nu} = \bra{\phi' _ \mu} \hat A \ket{\phi' _ \nu} = \bra{\phi _ \mu} \hat U ^ \dagger \hat A \hat U \ket{\phi _ \nu} =
	\bm U ^ \dagger \bm S^{-1} \bm A \bm S ^ {-1} \bm U 
\end{align}
while the contravariant representation transforms as
\begin{align}
	\hat{A} = \sum _ {\mu, \nu} \ket{\phi' _ \mu} (\bm A') _ {\mu \nu} \bra{\phi' _ \nu} = 
	\sum _ {\mu, \nu} \hat U \ket{\phi _ \mu} (\bm A') _ {\mu \nu} \bra{\phi _ \nu} \hat U ^ \dagger = 
	\sum _ {\mu, \nu} (\bm S^{-1} \bm U \bm A' \bm U^\dagger \bm S^{-1}) _ {\mu \nu} \ket{\phi _ \mu}\bra{\phi _ \nu}
\end{align}
which can be rewritten as
\begin{align}
	(\bm A') _ {\mu \nu} = ( \bm S ^ {-1} \bm U ^ \dagger \bm A \bm U \bm S ^ {-1}) _ {\mu \nu} \:.
\end{align}
Here, we used $\bra{\phi _ \mu} \hat U ^ {(\dagger)} \ket{\phi _ \nu} = (\bm U ^ {(\dagger)}) _ {\mu \nu}$ and unitarity condition $\bm U \bm S^ {-1} \bm U ^ \dagger = \bm S$.

The covariant representation transforms with $\bm S^{-1}$ directly next to operator components, which is the same as in
the completeness relation, while the contravariant representation transforms with $\bm U/\bm U^\dagger$ directly next to components. We note that in an orthonormal basis, overlap matrices become identity matrices, and hence the
representations become identical.

\subsection{Mixed Operator Representation in CP2K}

In CP2K, most operators are represented covariantly, while the density matrix is represented contravariantly.
This has an important effect on the numerical calculation of the trace of operators - in the orthonormal
basis, expectation values of operators are often evaluated as
\begin{align}
	\left \langle \hat A \right \rangle = \Tr{\hat \rho \hat A} = \sum _ m \bra{\psi _ m} \hat \rho \hat A \ket{\psi _ m}
\end{align}
Expressing the orthonormal vectors in terms of the non-orthogonal basis leads to
\begin{align}
	\Tr{\hat \rho \hat A} = 
	\sum _ {\mu, \nu, m} \bra{\phi _ \mu} C ^ * _ {\mu m} \hat \rho \hat A C _ {\nu m} \ket{\phi _ \nu} =
	\sum _ {\mu, \nu, \mu', \nu', m} C _ {\nu m} C^* _ {\mu m} S _ {\mu \mu'} \rho _ {\mu' \nu'} \bra{\phi _ {\nu'}} \hat A \ket{\phi _ \nu} \:,
\end{align}
where we substituted for the contravariant representation of $\hat \rho$. We now substitute from \eqref{cc_sinv},
which leads to
\begin{align}
	\Tr{\hat \rho \hat A} = \sum _ {\nu, \nu'} \rho _ {\nu \nu'} \bra{\phi _ {\nu'}} \hat A \ket{\phi _ {\nu}} \:,
\end{align}
meaning that using the covariant representation for $\hat A$ will lead to operator trace being calculated
as numerical trace of the matrix representation.

\subsection{Propagation of Matrices in Overlapping Basis}

The equation of motion ((1) in the main text) can be
represented in the overlapping Gaussian basis using the contravariant representation
for the density operator and the covariant representation for the
Hamiltonian operator as
\begin{align}
	\bm S \frac{\partial \bm \rho}{\partial t} \bm S = - \frac{i}{\hbar} \left ( \bm H(t) \bm \rho(t) \bm S - \bm S \bm \rho(t) \bm H(t) \right )
\end{align}
\begin{align}
	\frac{\partial \bm \rho}{\partial t} = - \frac{i}{\hbar} \left ( \bm S ^ {-1} \bm H(t) \bm \rho(t) - \bm \rho(t) \bm H(t) \bm S ^ {-1} \right )
\end{align}

An approximate solution to this differential equation would predict a density matrix at time $t + \Delta t$ using density matrix
at time $t$ as
\begin{align}
	\bm \rho (t + \Delta t) = e ^ {- (i/\hbar) \bm S ^ {-1} \bm H(t) \Delta t} \bm \rho(t) e ^ {(i/\hbar) \bm H(t) \bm S^ {-1} \Delta t}
	\label{no_etrs}
\end{align}
i.e. by neglecting the variation of $\bm H$ via approximation $\forall t' \in [t, t+\Delta t]: \bm H(t') = \bm H(t)$.

The variation of $\bm H$ with time can be included by self-consistent schemes such as the
enforced time reversal symmetry scheme (ETRS), as described in the main text.

\section{Exponentiation Methods}
\label{supp:exp_methods}

\subsection{Exact Diagonalisation}

In the exact diagonalisation scheme, Hamiltonian matrix $\bm H (t)$ is diagonalized at each time $t$, which
results in eigenvectors $\bm C (t)$
\begin{align}
	\bm H (t) \bm C (t) = \bm S \bm C(t) \bm \Lambda (t)
\end{align}
where $\bm \Lambda (t)$ is the diagonal matrix of eigenvalues. Using \eqref{cc_sinv}
\begin{align}
	\bm H(t) = \bm S \bm C(t) \bm \Lambda (t) \bm C ^ \dagger (t) \bm S
\end{align}
and therefore
\begin{align}
	e ^ {- (i/\hbar) \Delta t \bm S ^ {-1} \bm H(t)} &= \sum _ {n = 0} ^ \infty \left ( \frac{-i \Delta t}{\hbar} \right ) ^ n \frac{1}{n!}
	( \bm C(t) \bm \Lambda (t) \bm C^\dagger (t) \bm S ) ^ n = \\
	&= \sum _ {n = 0} ^ \infty \frac{1}{n!} \bm C(t) \left ( \frac{-i \Delta t}{\hbar} \bm \Lambda (t) \right ) ^ n \bm C ^ \dagger (t) \bm S
	= \bm C(t) e ^ {- (i/\hbar) \Delta t \bm \Lambda(t)} \bm C ^ \dagger (t) \bm S
\end{align}

where the exponential of diagonal matrix $-(i/\hbar) \Delta t \bm \Lambda (t)$ is trivial to evaluate.

\subsection{BCH Scheme}

Baker-Campbell-Hausdorff formula\cite{PropagatorsCastroRubio,PropagatorsORourke} establishes that for two operators $\hat A$ and $\hat B$, the
product
\begin{align}
	e ^ {\hat A} \hat B e ^ {- \hat A} = \sum _ {n = 0} ^ \infty \frac{1}{n!} [\hat A, \hat B] _ n
\end{align}
where
\begin{align}
	[\hat A, \hat B] _ n = [\hat A, [\hat A, \hat B] _ {n-1}], \: [\hat A, \hat B] _ 0 = \hat B
\end{align}

In non-orthogonal basis, the commutators are resolved as
\begin{align}
	([\hat A, \hat B]_n)_{\mu \nu} &= (\bra{\phi _ \mu} \hat A [\hat A, \hat B] _ {n-1} \ket{\phi _ \nu} - 
	\bra{\phi _ \mu} [\hat A, \hat B] _ {n-1} \hat A \ket{\phi _ \nu}) = \\
	&= \sum _ {\mu', \nu'} \left ( A _ {\mu \mu'} S ^ {-1} _ {\mu' \nu'} ([\hat A, \hat B]_{n-1}) _ {\nu' \nu}-
	([\hat A, \hat B]_{n-1})_{\mu \mu'} S ^ {-1} _ {\mu' \nu'} A _ {\nu' \nu}\right )
\end{align}
where
\begin{align}
	([\hat A, \hat B]_n) _ {\mu \nu} = \bra{\phi _ \mu} [\hat A, \hat B] _ n \ket{\phi _ \mu}
\end{align}
The mathematics is again simplified if we choose a mixed representation - we shall represent $\hat A$ as covariant
and the commutators as contravariant, i.e. we define $C _ {n, \mu \nu}$
\begin{align}
	[\hat A, \hat B]_n = \sum _ {\mu, \nu} C_{n,\mu \nu} \ket{\phi _ \mu} \bra{\phi _ \nu}
\end{align}
\begin{align}
	([\hat A, \hat B]_n)_{\mu \nu} = (\bm S \bm C _ {n} \bm S) _ {\mu \nu}
\end{align}

For the simple propagation scheme, we can identify $\hat A = -(i/\hbar) \Delta t \hat H(t)$ and in the atomic
orbital basis
\begin{align}
	\bm S \bm \rho (t + \Delta t) \bm S = \sum _ {n = 0} ^ \infty \frac{1}{n!} \bm S \bm C _ n \bm S
\end{align}
\begin{align}
	\bm \rho (t + \Delta t) = \sum _ {n = 0} ^ \infty \frac{1}{n!} \bm C _ n
\end{align}
with
\begin{align}
	\bm S \bm C _ n \bm S = \frac{-i}{\hbar} \Delta t \left ( \bm H(t) \bm C _ {n-1} \bm S - \bm S \bm C _ {n-1} \bm H(t)\right )
\end{align}
\begin{align}
	\bm C _ n = \frac{-i}{\hbar} \Delta t \left ( \bm S ^ {-1} \bm H(t) \bm C _ {n-1} - \bm C _ {n-1} \bm H(t) \bm S ^ {-1}\right )
\end{align}
and with
\begin{align}
	\bm C _ 0 = \bm \rho (t)
\end{align}

More specifically, one would typically define $\bm A = \frac{-i}{\hbar} \Delta t \bm S ^ {-1} \bm H (t)$ and then write
\begin{align}
    \bm C _ n = \bm A \bm C_{n-1} + \bm C _ {n-1} \bm A ^ \dagger
\end{align}
since both $\bm H(t)$ and $\bm S ^ {-1}$ are Hermitian.

In practice, we need to introduce a convergence parameter $\epsilon$ which truncates the series at smallest $n$ for which $||\bm C_n|| < \epsilon$,
assuming uniform convergence.

\section{Conservation of Idempotency}
\label{supp:idempotency}

As long as the effective Hamiltonian is hermitian, the density operator is idempotent at all times
if it is idempotent initially. For generally non-hermitian Hamiltonian, the equation of motion for the density operator would read
\begin{align}
    \frac{\partial \hat{\rho}}{\partial t} = \frac{-i}{\hbar} \left (
        \hat H (t) \hat{\rho}(t) - \hat{\rho}(t) \hat{H} ^ \dagger (t)
    \right )
\end{align}
From this, we can construct equation of motion for $\hat{\rho}^2 (t)$
\begin{align}
    \hat{\rho} (t) \frac{\partial \rho}{\partial t} + \frac{\partial \rho}{\partial t} \hat{\rho} (t)= - \frac{i}{\hbar} \left (
       \hat{\rho} (t) \hat H (t) \hat \rho (t) - \hat{\rho} ^ 2 (t) \hat{H}^\dagger(t) +
       \hat{H}(t) \hat{\rho} ^ 2 (t) - \hat{\rho}(t) \hat{H}^\dagger(t) \hat{\rho}(t)
    \right )
\end{align}
\begin{align}
    \frac{\partial (\hat{\rho} ^ 2)}{\partial t} = - \frac{i}{\hbar} \left (
        \hat H (t) \hat{\rho} ^ 2 (t) - \hat{\rho} ^ 2 (t) \hat{H}^\dagger(t)
    \right ) - \frac{i}{\hbar} \hat{\rho}(t) \left ( \hat H(t) - \hat{H}^\dagger (t)\right ) \hat{\rho}(t)
\end{align}

i.e. if at time $t$, $\hat{\rho} ^ 2(t) = \hat{\rho}(t)$, then such condition is true at all times,
as the equation of motion is the same for $\hat{\rho}(t)$ and $\hat{\rho}^2(t)$ at every instant,
as long as the Hamiltonian is hermitian.

In the overlapping basis, the form of idempotency condition is given by overlapping resolution of identity \eqref{resId_overlap}
\begin{align}
    \bra{\phi _ \mu} \hat \rho \ket{\phi _ \nu} &= (\bm S \bm \rho \bm S) _ {\mu \nu} =
    \bra{\phi _ \mu} \hat \rho ^ 2 \ket{\phi _ \nu} = \nonumber \\
    &= \sum _ {\mu', \nu'}
    \bra{\phi _ \mu} \hat \rho \ket{\phi _ {\mu'}} S _ {\mu' \nu'} \bra{\phi _ {\nu'}}\ket{\phi _ \nu} = (\bm S \bm \rho \bm S \bm S ^ {-1} \bm S \bm \rho \bm S) _ {\mu \nu}
\end{align}
\begin{align}
    \bm \rho = \bm \rho \bm S \bm \rho \label{idempotence_overlap}
\end{align}

In the code, we track the idempotency deviation by outputting
\begin{align}
    \Tr{\hat \rho ^ 2} - \Tr{\hat \rho} = \sum _ \mu (\bm \rho \bm S \bm \rho \bm S) _ {\mu \mu} - 
    (\bm \rho \bm S) _ {\mu \mu} = \sum _ \mu (\bm \rho \bm S (\bm \rho \bm S - \bm I)) _ {\mu \mu}
\end{align}

\section{Hartree and COHSEX Terms in the Gaussian overlapping basis}
\label{supp:hartree_and_cohsex}

\subsection{Hartree Term}

In the position basis, Hartree term is given as
\begin{align}
	\bra{\bm r} \hat V ^ \text{Hartree} (t)\ket{\bm r'} = V^\text{Hartree}(\bm r, \bm r', t) =
	\delta(\bm r - \bm r') \int d^3 r'' \frac{e^2\rho(\bm r'', t)}{4 \pi \epsilon _ 0|\bm r - \bm r''|}
\end{align}
where $\rho(\bm r', t) = \bra{\bm r'} \hat \rho(t) \ket{\bm r'}$. Substituting for covariant
representation of the Hartree term in the atomic orbital basis and contravariant representation
of the density matrix in the atomic orbital basis, we obtain
\begin{align}
	\sum _ {\mu, \nu, \mu', \nu'} \phi _ \mu (\bm r) S ^ {-1} _ {\mu \nu} V ^ \text{Hartree} _ {\nu \mu'} (t) S ^ {-1} _ {\mu' \nu'} \phi ^ * _ {\nu'} (\bm r') =
	\braket{\bm r}{\bm r'} \int d ^ 3 r'' \frac{e^2}{4 \pi \epsilon_0|\bm r - \bm r''|} \sum _ {\mu, \nu} \phi _ \mu (\bm r'') \phi _ \nu ^ * (\bm r'') \rho _ {\mu \nu} (t)
\end{align}
Multiplying by $\braket{\phi_\lambda}{\bm r} \braket{\bm r'}{\phi _ \sigma}$ and integrating in $\bm r$ and $\bm r'$,
we obtain
\begin{align}
	V ^ \text{Hartree} _ {\lambda \sigma} (t) = \int d ^ 3 r d ^ 3 r' d ^ 3 r'' \braket{\phi _ \lambda}{\bm r} \braket{\bm r}{\bm r'} \braket{\bm r'}{\phi _ \sigma}
	\frac{e^2}{4 \pi \epsilon_0|\bm r - \bm r''|} \sum _ {\mu \nu} \phi _ \mu (\bm r'') \phi _ \nu ^ * (\bm r'')
\end{align}
Resolving the completeness relation in $\bm r'$ leads to\cite{RTBSEAttaccalite}
\begin{align}
	V ^ \text{Hartree} _ {\lambda\sigma} (t) = \sum _ {\mu, \nu}
	\int d ^ 3 r d ^ 3 r'' \phi ^ * _ \lambda (\bm r) \phi _ \sigma (\bm r) \frac{e^2}{4 \pi \epsilon _ 0|\bm r - \bm r''|} \phi ^ * _ {\nu} (\bm r'') \phi _ \mu (\bm r'')
	\rho _ {\mu \nu} (t) = \sum _ {\mu, \nu} (\lambda \sigma | \nu \mu) \rho _ {\mu \nu} (t)
\end{align}
For real orbitals then also $V^\text{Hartree} _ {\lambda \sigma} = \sum _ {\mu \nu} (\lambda \sigma | \mu \nu) \rho _ {\mu \nu}$.

\subsection{COHSEX Term}

The Coulomb hole term has the following position base representation\cite{RTBSEAttaccalite}
\begin{align}
	\bra{\bm r} \hat \Sigma ^ \text{COH} (t) \ket{\bm r'} = - \frac{1}{2} W(\bm r, \bm r', \omega = 0) \delta(\bm r - \bm r')
\end{align}
The screened exchange is represented as\cite{RTBSEAttaccalite}
\begin{align}
	\bra{\bm r} \hat \Sigma ^ \text{SEX} (t) \ket{\bm r'} = i \hbar W(\bm r, \bm r', \omega = 0) G ^ < (\bm r, \bm r', t)
\end{align}
with
\begin{align}
	\hat{\Sigma} ^ \text{COHSEX} (t) = \hat \Sigma ^ \text{COH} (t) + \hat \Sigma ^ \text{SEX} (t)
\end{align}
where $G^<(\bm r, \bm r', t)$ is the time-diagonal lesser Green's function\cite{RTBSEAttaccalite}.

Under the approximation that the screened exchange $W$ does not change as time passes, the
Coulomb hole part of the self-energy is constant, and hence does not contribute to the dynamics\cite{RTBSEAttaccalite}.

The screened exchange part can be expressed in terms of the density matrix, since
\begin{align}
	G^<(\bm r, \bm r', t) = \frac{i}{\hbar} \rho(\bm r, \bm r', t)
\end{align}
leading to
\begin{align}
	\sum _ {\mu, \nu, \mu', \nu'} \phi _ \mu (\bm r) S ^ {-1} _ {\mu \nu} \Sigma ^ \text{SEX} _ {\nu \mu'} S ^ {-1} _ {\mu' \nu'} \phi _ {\nu'} ^ * (\bm r') =
	- W(\bm r, \bm r', \omega = 0) \bra{\bm r} \hat \rho(t) \ket{\bm r'}
\end{align}
Similarly as before, multiplying by $\braket{\phi_\lambda}{\bm r} \braket{\bm r'}{\phi _ \sigma}$ and integrating in $\bm r$ and $\bm r'$,\cite{RTBSEAttaccalite}
\begin{align}
	\Sigma ^ \text{SEX} _ {\lambda \sigma}(t) = - \int d ^ 3 r \int d^ 3 r' \sum _ {\mu, \nu} \phi _ \lambda ^ * (\bm r) \phi _ \mu (\bm r) W(\bm r, \bm r', \omega = 0)
	\phi _ \sigma(\bm r') \phi _ \nu ^ * (\bm r') \rho _ {\mu \nu} (t)
\end{align}
\begin{align}
	\Sigma ^ \text{SEX} _ {\lambda \sigma} (t) = - \sum _ {\mu,\nu} W_{\lambda \mu, \nu \sigma} \rho _ {\mu \nu} (t)
\end{align}
where we used the $W_{\lambda \mu, \nu \sigma}$ notation for the screened four-center integral.

\section{RI Approximation}
\label{supp:ri}

Resolution of identity approximation can be used to reduce the scaling of electron repulsion integrals
with number of basis functions. The basic idea is that the basis pair products are
substituted by summations over auxiliary basis functions which minimize difference from true ERIs, specifically\cite{LSGWGramlWilhelm}
\begin{align}
	\phi _ \mu (\bm r) \phi _ \nu (\bm r) = \sum _ P B ^ P _ {\mu \nu} \phi _ P (\bm r)
\end{align}
where $B ^ P _ {\mu \nu} = \sum _ Q m _ {PQ} ^ {-1} (Q ][ \mu \nu)$, where $(Q ][ \mu \nu)$ is
a 3-center integral with $m(\bm r, \bm r')$ as the metric.

This simplifies the calculation of Hartree term to
\begin{align}
	V^\text{Hartree} _ {\lambda \sigma} = \sum _ {P,Q} (\lambda \sigma ][ P) (\bm m ^ {-1} \bm V ^\text{H,A} \bm m ^ {-1}) _ {PQ} \sum _ {\mu,\nu} (Q ][ \nu \mu ) \rho _ {\mu \nu} (t)
\end{align}
where
\begin{align}
	V^\text{H,A} _ {PQ} = \int d^3 r d ^ 3 r' \phi ^ * _ P (\bm r) \frac{1}{|\bm r - \bm r'|} \phi _ Q (\bm r')
\end{align}

Similar approximation can be made for the screened exchange self-energy, with
\begin{align}
	\Sigma ^ \text{SEX} _ {\lambda \sigma} (t) = -\frac{1}{\hbar} \sum _ {P,Q,\mu,\nu} (\lambda \mu ][ P) (\bm m ^ {-1} \bm W ^ \text{A} \bm m ^ {-1}) _ {PQ} (Q ][ \nu \sigma ) \rho _ {\mu \nu} (t) \label{sigma_ri}
\end{align}

\section{RI Auxiliary Basis Set Convergence}
\label{supp:ri_convergence}

We varied the RI auxiliary basis set for a fixed orbital basis set (aug-cc-pVDZ), specifically,
we used aug-cc-pVDZ-RIFIT (DZ), aug-cc-pVTZ-RIFIT (TZ), aug-cc-pVQZ-RIFIT (QZ) and aug-cc-pV5Z-RIFIT (5Z).
The recorded LUMO-HOMO (lowest unoccupied/highest occupied molecular orbital) gap,
as recovered from the $GW$, are shown below in
\figref{fig:ri_convergence}, with respect to the gap observed for 5Z calculation.

\begin{figure}[h]
	\centering
	\resizebox{0.7\columnwidth}{!}{\input{graphics/ri_convergence.pgf}}
	\caption{LUMO-HOMO gap of molecules in the Thiel's set with differing auxiliary basis set.
	The variation with respect to 5Z is around 10 meV, going up to about 30 meV for largest deviation.}
	\label{fig:ri_convergence}
\end{figure}

\section{RI Cutoff Radius Convergence}
\label{supp:ri_cutoff_convergence}

\begin{figure}[h!]
	\centering
	\resizebox{0.68\columnwidth}{!}{\input{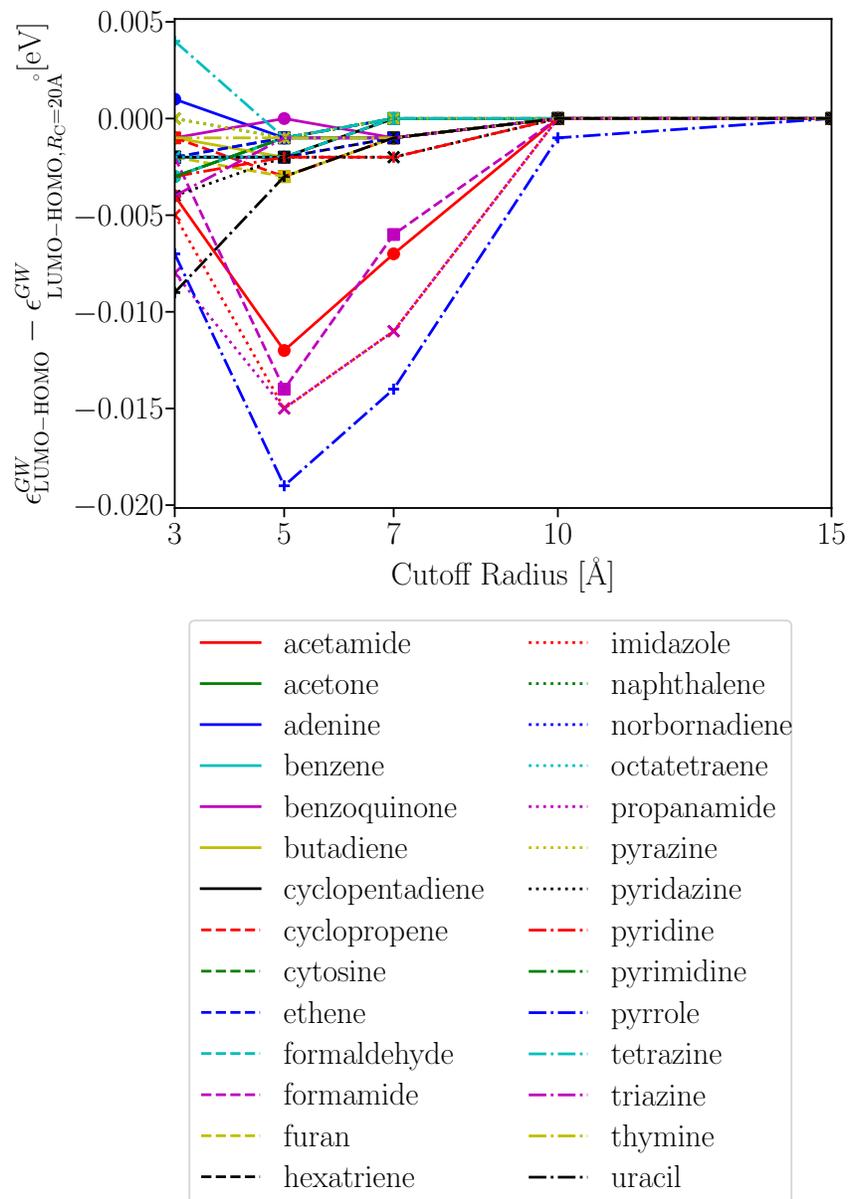}}
	\caption{LUMO-HOMO gap of molecules in the Thiel's set with different RI cutoff radius.
	The RI basis used here was the aug-cc-pvtz-rifit. The values are compared to a
	gap with 20~$\mathrm{\AA}$ cutoff radius. The values start to converge from 7~$\mathrm{\AA}$ used in the main text.}
	\label{fig:ri_cutoff_convergence}
\end{figure}

We varied the cutoff-radius of the RI-metric used for the 3-centre integrals in RI approximation to study the effect on the LUMO-HOMO gap, shown
\figref{fig:ri_cutoff_convergence}.

\FloatBarrier

\section{Timestep Convergence in HHG of Cysteine}
\label{supp:timestep_convergence}

\begin{figure}[ht]
	\centering
	\resizebox{0.49\columnwidth}{!}{\input{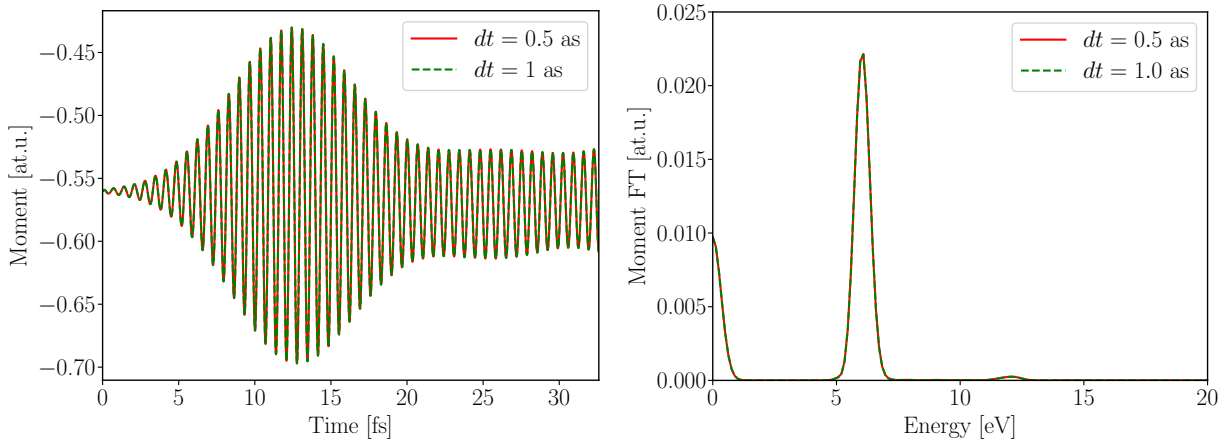}}
	\resizebox{0.49\columnwidth}{!}{\input{graphics/cysteine_ft_convergence.pgf}}
	\caption{On the left, the oscillating dipole moment series in cysteine for different
	size of the timestep in the simulation. No visible departure within the investigated times
	is observed. On the right, Fourier transform of the series using Gaussian window function
	centered at 22.6~fs with 2.0~fs spread. Again, no visible difference is observed between the
	two spectra.}
	\label{fig:cysteine_convergence}
\end{figure}

We checked a smaller timestep of 0.5~as, which showed no divergence from the 1~as timestep used in the
main body of the paper. The time series of the oscillating dipole moment as well as the Fourier transform
using Gaussian window function are shown in \figref{fig:cysteine_convergence}.

Since no visible difference is present in either the time series or in the spectra, we conclude that the
simulation is converged in the timestep size.

\section{Computational Scaling}
\label{supp:scaling}

\begin{figure}[h]
	\centering
	\includegraphics[width=0.49\columnwidth]{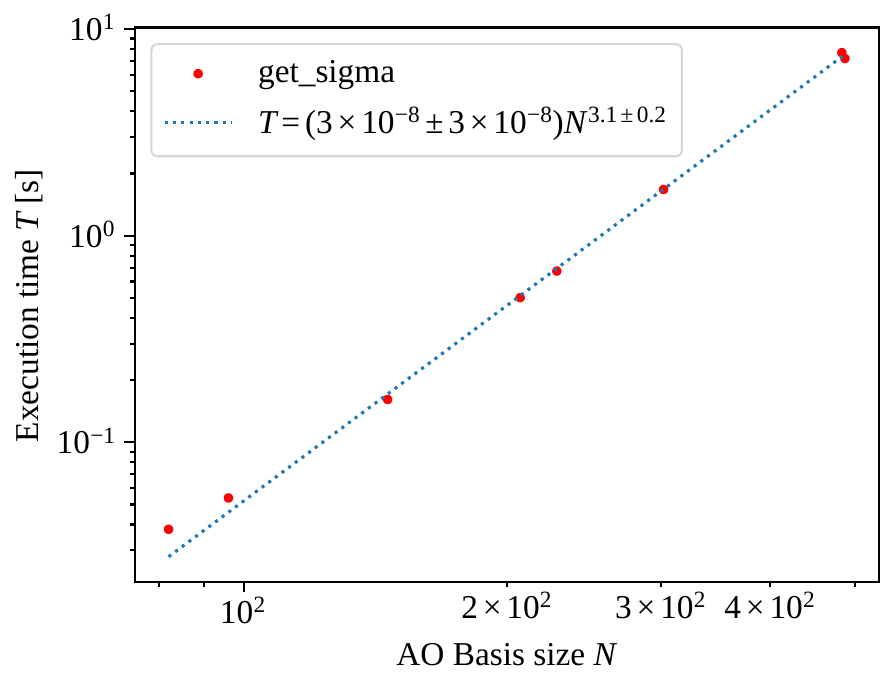}\hfill
	\includegraphics[width=0.49\columnwidth]{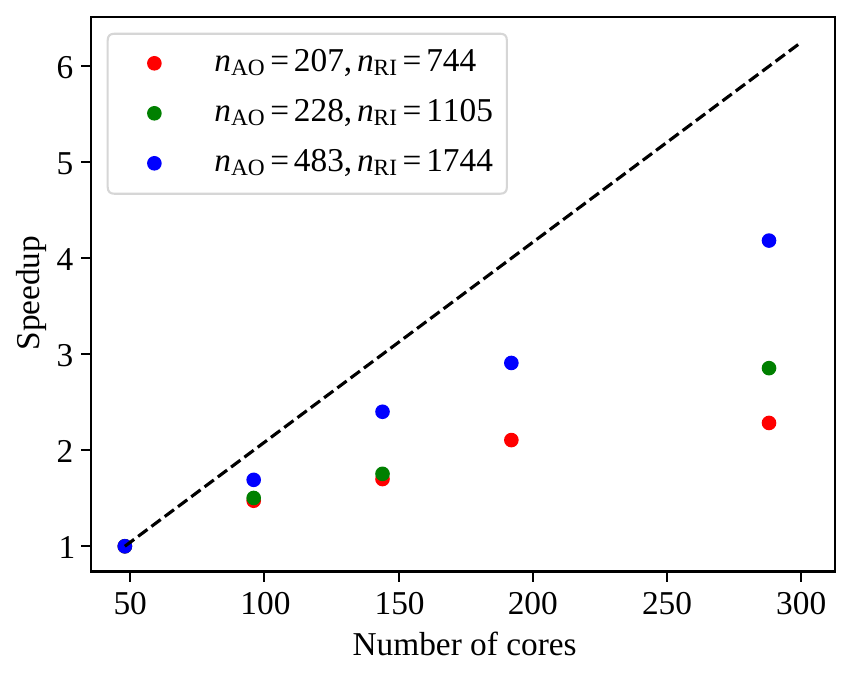}
	\caption{Left: Scaling of the execution time of the most expensive routine in the calculation. We observe
	approximately cubic scaling with the size of the basis. All data use 16 MPI ranks, each with
	3 OMP threads. Right: Strong scaling investigation -- for a fixed basis set size, we increase the number of
	cores and note the speedup. We see that the speedup plateaus later for larger systems, hinting at possible presence of
	weak scaling phenomenon. The dashed line shows perfect parallelisation speedup.}
	\label{fig:basis_scaling}
\end{figure}

We investigated the scaling of the execution time of the most expensive routine \verb+get_sigma+ (determination
of the screened exchange self energy) of the
code as a function of the basis set size. The routine is called many times during the execution of the code --
we report the average timing per one call. Typically, the routine is called about 2-5 times per timestep and
the propagation in the validation runs used 20~000 time steps. We observed a $T \propto N^{3.1 \pm 0.2}$ dependence, as shown
in \figref{fig:basis_scaling}.

The $N^3$ scaling is slower than anticipated from \eqref{sigma_ri}, where the most expensive steps are
expected to have $N^4$ scaling (for example contraction of $\sum _ \mu (Q][\nu \mu) \rho _ {\mu \nu}$).
The reduced scaling shows that the parallel efficiency may not yet be fully realized for the number of cores
used in the test.

\begin{figure}[h]
	\centering
	\includegraphics[width=0.7\columnwidth]{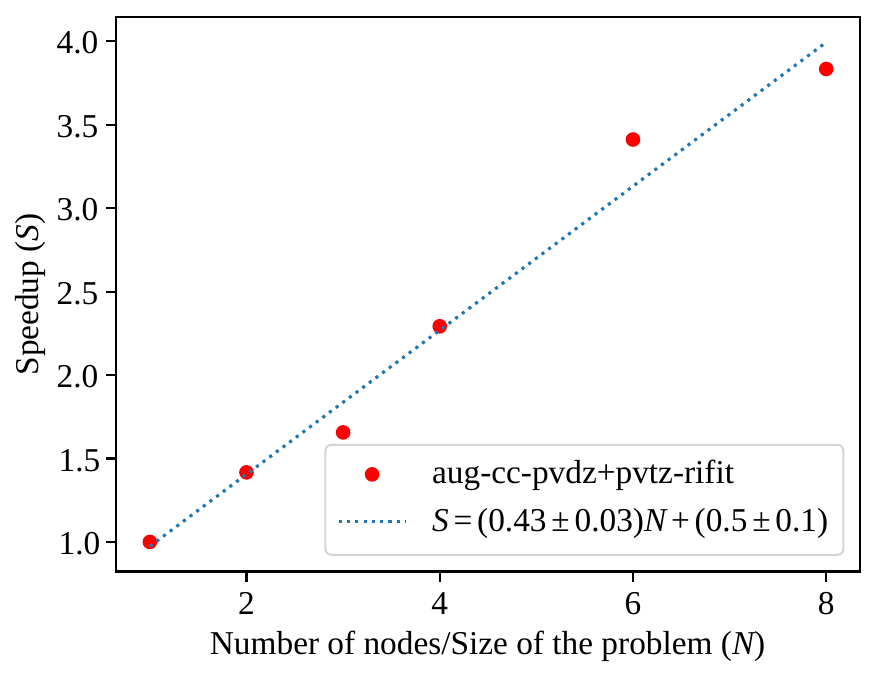}
	\caption{The speedup obtained for system size increasing approximately together with
	the number of cores used for the calculation. The approximately linear dependence is consistent
	with Gustafson's law\cite{ScalingGustafson}.}
	\label{fig:weak_scaling}
\end{figure}

The dependence of the maximum attainable speedup on the system size shows the possibility of weak scaling phenomenon,
which we decided to further validate. We ran a series of calculations where the computational effort
(cube of the basis set size, $N ^ 3$) increased approximately along with the number of resources
used for the calculation (number of computational nodes, each with 48 processors). This produced
series of average times required per call of \verb+get_sigma+, $T_N(N)$. We also calculated
the average time required per call of the \verb+get_sigma+ routine when running on a single node
for all of these cases, $T_N(1)$. The chosen parameters are shown in Table \ref{tab:weak_scaling},
while the speedup $T_N(1)/T_N(N)$ is shown in \figref{fig:weak_scaling}.

\begin{table}[h]
	\begin{tabular}{c|c|c}
		\# Nodes & $N$ & $N^3$ \\
		\hline
		1 & 146 & $3.11 \times 10 ^ 6$ \\
		2 & 183 & $6.13 \times 10 ^ 6$ \\
		3 & 210 & $9.26 \times 10 ^ 6$ \\
		4 & 228 & $11.85 \times 10^6$ \\
		6 & 261 & $17.78 \times 10^6$ \\
		8 & 302 & $27.54 \times 10^6$ \\
	\end{tabular}
	\caption{The number of nodes and the approximately proportional problem size used for the
	demonstration of Gustafson's law in \figref{fig:weak_scaling}.}
	\label{tab:weak_scaling}
\end{table}

\FloatBarrier

%% file: graphics/cysteine_ft_convergence.pgf
\begingroup%
\makeatletter%
\begin{pgfpicture}%
\pgfpathrectangle{\pgfpointorigin}{\pgfqpoint{6.100111in}{4.545222in}}%
\pgfusepath{use as bounding box, clip}%
\begin{pgfscope}%
\pgfsetbuttcap%
\pgfsetmiterjoin%
\definecolor{currentfill}{rgb}{1.000000,1.000000,1.000000}%
\pgfsetfillcolor{currentfill}%
\pgfsetlinewidth{0.000000pt}%
\definecolor{currentstroke}{rgb}{1.000000,1.000000,1.000000}%
\pgfsetstrokecolor{currentstroke}%
\pgfsetdash{}{0pt}%
\pgfpathmoveto{\pgfqpoint{0.000000in}{0.000000in}}%
\pgfpathlineto{\pgfqpoint{6.100111in}{0.000000in}}%
\pgfpathlineto{\pgfqpoint{6.100111in}{4.545222in}}%
\pgfpathlineto{\pgfqpoint{0.000000in}{4.545222in}}%
\pgfpathlineto{\pgfqpoint{0.000000in}{0.000000in}}%
\pgfpathclose%
\pgfusepath{fill}%
\end{pgfscope}%
\begin{pgfscope}%
\pgfsetbuttcap%
\pgfsetmiterjoin%
\definecolor{currentfill}{rgb}{1.000000,1.000000,1.000000}%
\pgfsetfillcolor{currentfill}%
\pgfsetlinewidth{0.000000pt}%
\definecolor{currentstroke}{rgb}{0.000000,0.000000,0.000000}%
\pgfsetstrokecolor{currentstroke}%
\pgfsetstrokeopacity{0.000000}%
\pgfsetdash{}{0pt}%
\pgfpathmoveto{\pgfqpoint{0.938111in}{0.672111in}}%
\pgfpathlineto{\pgfqpoint{5.898111in}{0.672111in}}%
\pgfpathlineto{\pgfqpoint{5.898111in}{4.368111in}}%
\pgfpathlineto{\pgfqpoint{0.938111in}{4.368111in}}%
\pgfpathlineto{\pgfqpoint{0.938111in}{0.672111in}}%
\pgfpathclose%
\pgfusepath{fill}%
\end{pgfscope}%
\begin{pgfscope}%
\pgfsetbuttcap%
\pgfsetroundjoin%
\definecolor{currentfill}{rgb}{0.000000,0.000000,0.000000}%
\pgfsetfillcolor{currentfill}%
\pgfsetlinewidth{0.803000pt}%
\definecolor{currentstroke}{rgb}{0.000000,0.000000,0.000000}%
\pgfsetstrokecolor{currentstroke}%
\pgfsetdash{}{0pt}%
\pgfsys@defobject{currentmarker}{\pgfqpoint{0.000000in}{-0.048611in}}{\pgfqpoint{0.000000in}{0.000000in}}{%
\pgfpathmoveto{\pgfqpoint{0.000000in}{0.000000in}}%
\pgfpathlineto{\pgfqpoint{0.000000in}{-0.048611in}}%
\pgfusepath{stroke,fill}%
}%
\begin{pgfscope}%
\pgfsys@transformshift{0.938111in}{0.672111in}%
\pgfsys@useobject{currentmarker}{}%
\end{pgfscope}%
\end{pgfscope}%
\begin{pgfscope}%
\definecolor{textcolor}{rgb}{0.000000,0.000000,0.000000}%
\pgfsetstrokecolor{textcolor}%
\pgfsetfillcolor{textcolor}%
\pgftext[x=0.938111in,y=0.574888in,,top]{\color{textcolor}{\fontfamily{\familydefault}\fontsize{16.000000}{19.200000}\selectfont\catcode`\^=\active\def^{\ifmmode\sp\else\^{}\fi}\catcode`\%=\active\def
\end{pgfscope}%
\begin{pgfscope}%
\pgfsetbuttcap%
\pgfsetroundjoin%
\definecolor{currentfill}{rgb}{0.000000,0.000000,0.000000}%
\pgfsetfillcolor{currentfill}%
\pgfsetlinewidth{0.803000pt}%
\definecolor{currentstroke}{rgb}{0.000000,0.000000,0.000000}%
\pgfsetstrokecolor{currentstroke}%
\pgfsetdash{}{0pt}%
\pgfsys@defobject{currentmarker}{\pgfqpoint{0.000000in}{-0.048611in}}{\pgfqpoint{0.000000in}{0.000000in}}{%
\pgfpathmoveto{\pgfqpoint{0.000000in}{0.000000in}}%
\pgfpathlineto{\pgfqpoint{0.000000in}{-0.048611in}}%
\pgfusepath{stroke,fill}%
}%
\begin{pgfscope}%
\pgfsys@transformshift{2.178111in}{0.672111in}%
\pgfsys@useobject{currentmarker}{}%
\end{pgfscope}%
\end{pgfscope}%
\begin{pgfscope}%
\definecolor{textcolor}{rgb}{0.000000,0.000000,0.000000}%
\pgfsetstrokecolor{textcolor}%
\pgfsetfillcolor{textcolor}%
\pgftext[x=2.178111in,y=0.574888in,,top]{\color{textcolor}{\fontfamily{\familydefault}\fontsize{16.000000}{19.200000}\selectfont\catcode`\^=\active\def^{\ifmmode\sp\else\^{}\fi}\catcode`\%=\active\def
\end{pgfscope}%
\begin{pgfscope}%
\pgfsetbuttcap%
\pgfsetroundjoin%
\definecolor{currentfill}{rgb}{0.000000,0.000000,0.000000}%
\pgfsetfillcolor{currentfill}%
\pgfsetlinewidth{0.803000pt}%
\definecolor{currentstroke}{rgb}{0.000000,0.000000,0.000000}%
\pgfsetstrokecolor{currentstroke}%
\pgfsetdash{}{0pt}%
\pgfsys@defobject{currentmarker}{\pgfqpoint{0.000000in}{-0.048611in}}{\pgfqpoint{0.000000in}{0.000000in}}{%
\pgfpathmoveto{\pgfqpoint{0.000000in}{0.000000in}}%
\pgfpathlineto{\pgfqpoint{0.000000in}{-0.048611in}}%
\pgfusepath{stroke,fill}%
}%
\begin{pgfscope}%
\pgfsys@transformshift{3.418111in}{0.672111in}%
\pgfsys@useobject{currentmarker}{}%
\end{pgfscope}%
\end{pgfscope}%
\begin{pgfscope}%
\definecolor{textcolor}{rgb}{0.000000,0.000000,0.000000}%
\pgfsetstrokecolor{textcolor}%
\pgfsetfillcolor{textcolor}%
\pgftext[x=3.418111in,y=0.574888in,,top]{\color{textcolor}{\fontfamily{\familydefault}\fontsize{16.000000}{19.200000}\selectfont\catcode`\^=\active\def^{\ifmmode\sp\else\^{}\fi}\catcode`\%=\active\def
\end{pgfscope}%
\begin{pgfscope}%
\pgfsetbuttcap%
\pgfsetroundjoin%
\definecolor{currentfill}{rgb}{0.000000,0.000000,0.000000}%
\pgfsetfillcolor{currentfill}%
\pgfsetlinewidth{0.803000pt}%
\definecolor{currentstroke}{rgb}{0.000000,0.000000,0.000000}%
\pgfsetstrokecolor{currentstroke}%
\pgfsetdash{}{0pt}%
\pgfsys@defobject{currentmarker}{\pgfqpoint{0.000000in}{-0.048611in}}{\pgfqpoint{0.000000in}{0.000000in}}{%
\pgfpathmoveto{\pgfqpoint{0.000000in}{0.000000in}}%
\pgfpathlineto{\pgfqpoint{0.000000in}{-0.048611in}}%
\pgfusepath{stroke,fill}%
}%
\begin{pgfscope}%
\pgfsys@transformshift{4.658111in}{0.672111in}%
\pgfsys@useobject{currentmarker}{}%
\end{pgfscope}%
\end{pgfscope}%
\begin{pgfscope}%
\definecolor{textcolor}{rgb}{0.000000,0.000000,0.000000}%
\pgfsetstrokecolor{textcolor}%
\pgfsetfillcolor{textcolor}%
\pgftext[x=4.658111in,y=0.574888in,,top]{\color{textcolor}{\fontfamily{\familydefault}\fontsize{16.000000}{19.200000}\selectfont\catcode`\^=\active\def^{\ifmmode\sp\else\^{}\fi}\catcode`\%=\active\def
\end{pgfscope}%
\begin{pgfscope}%
\pgfsetbuttcap%
\pgfsetroundjoin%
\definecolor{currentfill}{rgb}{0.000000,0.000000,0.000000}%
\pgfsetfillcolor{currentfill}%
\pgfsetlinewidth{0.803000pt}%
\definecolor{currentstroke}{rgb}{0.000000,0.000000,0.000000}%
\pgfsetstrokecolor{currentstroke}%
\pgfsetdash{}{0pt}%
\pgfsys@defobject{currentmarker}{\pgfqpoint{0.000000in}{-0.048611in}}{\pgfqpoint{0.000000in}{0.000000in}}{%
\pgfpathmoveto{\pgfqpoint{0.000000in}{0.000000in}}%
\pgfpathlineto{\pgfqpoint{0.000000in}{-0.048611in}}%
\pgfusepath{stroke,fill}%
}%
\begin{pgfscope}%
\pgfsys@transformshift{5.898111in}{0.672111in}%
\pgfsys@useobject{currentmarker}{}%
\end{pgfscope}%
\end{pgfscope}%
\begin{pgfscope}%
\definecolor{textcolor}{rgb}{0.000000,0.000000,0.000000}%
\pgfsetstrokecolor{textcolor}%
\pgfsetfillcolor{textcolor}%
\pgftext[x=5.898111in,y=0.574888in,,top]{\color{textcolor}{\fontfamily{\familydefault}\fontsize{16.000000}{19.200000}\selectfont\catcode`\^=\active\def^{\ifmmode\sp\else\^{}\fi}\catcode`\%=\active\def
\end{pgfscope}%
\begin{pgfscope}%
\definecolor{textcolor}{rgb}{0.000000,0.000000,0.000000}%
\pgfsetstrokecolor{textcolor}%
\pgfsetfillcolor{textcolor}%
\pgftext[x=3.418111in,y=0.321777in,,top]{\color{textcolor}{\fontfamily{\familydefault}\fontsize{16.000000}{19.200000}\selectfont\catcode`\^=\active\def^{\ifmmode\sp\else\^{}\fi}\catcode`\%=\active\def
\end{pgfscope}%
\begin{pgfscope}%
\pgfsetbuttcap%
\pgfsetroundjoin%
\definecolor{currentfill}{rgb}{0.000000,0.000000,0.000000}%
\pgfsetfillcolor{currentfill}%
\pgfsetlinewidth{0.803000pt}%
\definecolor{currentstroke}{rgb}{0.000000,0.000000,0.000000}%
\pgfsetstrokecolor{currentstroke}%
\pgfsetdash{}{0pt}%
\pgfsys@defobject{currentmarker}{\pgfqpoint{-0.048611in}{0.000000in}}{\pgfqpoint{-0.000000in}{0.000000in}}{%
\pgfpathmoveto{\pgfqpoint{-0.000000in}{0.000000in}}%
\pgfpathlineto{\pgfqpoint{-0.048611in}{0.000000in}}%
\pgfusepath{stroke,fill}%
}%
\begin{pgfscope}%
\pgfsys@transformshift{0.938111in}{0.672111in}%
\pgfsys@useobject{currentmarker}{}%
\end{pgfscope}%
\end{pgfscope}%
\begin{pgfscope}%
\definecolor{textcolor}{rgb}{0.000000,0.000000,0.000000}%
\pgfsetstrokecolor{textcolor}%
\pgfsetfillcolor{textcolor}%
\pgftext[x=0.377333in, y=0.594999in, left, base]{\color{textcolor}{\fontfamily{\familydefault}\fontsize{16.000000}{19.200000}\selectfont\catcode`\^=\active\def^{\ifmmode\sp\else\^{}\fi}\catcode`\%=\active\def
\end{pgfscope}%
\begin{pgfscope}%
\pgfsetbuttcap%
\pgfsetroundjoin%
\definecolor{currentfill}{rgb}{0.000000,0.000000,0.000000}%
\pgfsetfillcolor{currentfill}%
\pgfsetlinewidth{0.803000pt}%
\definecolor{currentstroke}{rgb}{0.000000,0.000000,0.000000}%
\pgfsetstrokecolor{currentstroke}%
\pgfsetdash{}{0pt}%
\pgfsys@defobject{currentmarker}{\pgfqpoint{-0.048611in}{0.000000in}}{\pgfqpoint{-0.000000in}{0.000000in}}{%
\pgfpathmoveto{\pgfqpoint{-0.000000in}{0.000000in}}%
\pgfpathlineto{\pgfqpoint{-0.048611in}{0.000000in}}%
\pgfusepath{stroke,fill}%
}%
\begin{pgfscope}%
\pgfsys@transformshift{0.938111in}{1.411311in}%
\pgfsys@useobject{currentmarker}{}%
\end{pgfscope}%
\end{pgfscope}%
\begin{pgfscope}%
\definecolor{textcolor}{rgb}{0.000000,0.000000,0.000000}%
\pgfsetstrokecolor{textcolor}%
\pgfsetfillcolor{textcolor}%
\pgftext[x=0.377333in, y=1.334199in, left, base]{\color{textcolor}{\fontfamily{\familydefault}\fontsize{16.000000}{19.200000}\selectfont\catcode`\^=\active\def^{\ifmmode\sp\else\^{}\fi}\catcode`\%=\active\def
\end{pgfscope}%
\begin{pgfscope}%
\pgfsetbuttcap%
\pgfsetroundjoin%
\definecolor{currentfill}{rgb}{0.000000,0.000000,0.000000}%
\pgfsetfillcolor{currentfill}%
\pgfsetlinewidth{0.803000pt}%
\definecolor{currentstroke}{rgb}{0.000000,0.000000,0.000000}%
\pgfsetstrokecolor{currentstroke}%
\pgfsetdash{}{0pt}%
\pgfsys@defobject{currentmarker}{\pgfqpoint{-0.048611in}{0.000000in}}{\pgfqpoint{-0.000000in}{0.000000in}}{%
\pgfpathmoveto{\pgfqpoint{-0.000000in}{0.000000in}}%
\pgfpathlineto{\pgfqpoint{-0.048611in}{0.000000in}}%
\pgfusepath{stroke,fill}%
}%
\begin{pgfscope}%
\pgfsys@transformshift{0.938111in}{2.150511in}%
\pgfsys@useobject{currentmarker}{}%
\end{pgfscope}%
\end{pgfscope}%
\begin{pgfscope}%
\definecolor{textcolor}{rgb}{0.000000,0.000000,0.000000}%
\pgfsetstrokecolor{textcolor}%
\pgfsetfillcolor{textcolor}%
\pgftext[x=0.377333in, y=2.073399in, left, base]{\color{textcolor}{\fontfamily{\familydefault}\fontsize{16.000000}{19.200000}\selectfont\catcode`\^=\active\def^{\ifmmode\sp\else\^{}\fi}\catcode`\%=\active\def
\end{pgfscope}%
\begin{pgfscope}%
\pgfsetbuttcap%
\pgfsetroundjoin%
\definecolor{currentfill}{rgb}{0.000000,0.000000,0.000000}%
\pgfsetfillcolor{currentfill}%
\pgfsetlinewidth{0.803000pt}%
\definecolor{currentstroke}{rgb}{0.000000,0.000000,0.000000}%
\pgfsetstrokecolor{currentstroke}%
\pgfsetdash{}{0pt}%
\pgfsys@defobject{currentmarker}{\pgfqpoint{-0.048611in}{0.000000in}}{\pgfqpoint{-0.000000in}{0.000000in}}{%
\pgfpathmoveto{\pgfqpoint{-0.000000in}{0.000000in}}%
\pgfpathlineto{\pgfqpoint{-0.048611in}{0.000000in}}%
\pgfusepath{stroke,fill}%
}%
\begin{pgfscope}%
\pgfsys@transformshift{0.938111in}{2.889711in}%
\pgfsys@useobject{currentmarker}{}%
\end{pgfscope}%
\end{pgfscope}%
\begin{pgfscope}%
\definecolor{textcolor}{rgb}{0.000000,0.000000,0.000000}%
\pgfsetstrokecolor{textcolor}%
\pgfsetfillcolor{textcolor}%
\pgftext[x=0.377333in, y=2.812599in, left, base]{\color{textcolor}{\fontfamily{\familydefault}\fontsize{16.000000}{19.200000}\selectfont\catcode`\^=\active\def^{\ifmmode\sp\else\^{}\fi}\catcode`\%=\active\def
\end{pgfscope}%
\begin{pgfscope}%
\pgfsetbuttcap%
\pgfsetroundjoin%
\definecolor{currentfill}{rgb}{0.000000,0.000000,0.000000}%
\pgfsetfillcolor{currentfill}%
\pgfsetlinewidth{0.803000pt}%
\definecolor{currentstroke}{rgb}{0.000000,0.000000,0.000000}%
\pgfsetstrokecolor{currentstroke}%
\pgfsetdash{}{0pt}%
\pgfsys@defobject{currentmarker}{\pgfqpoint{-0.048611in}{0.000000in}}{\pgfqpoint{-0.000000in}{0.000000in}}{%
\pgfpathmoveto{\pgfqpoint{-0.000000in}{0.000000in}}%
\pgfpathlineto{\pgfqpoint{-0.048611in}{0.000000in}}%
\pgfusepath{stroke,fill}%
}%
\begin{pgfscope}%
\pgfsys@transformshift{0.938111in}{3.628911in}%
\pgfsys@useobject{currentmarker}{}%
\end{pgfscope}%
\end{pgfscope}%
\begin{pgfscope}%
\definecolor{textcolor}{rgb}{0.000000,0.000000,0.000000}%
\pgfsetstrokecolor{textcolor}%
\pgfsetfillcolor{textcolor}%
\pgftext[x=0.377333in, y=3.551799in, left, base]{\color{textcolor}{\fontfamily{\familydefault}\fontsize{16.000000}{19.200000}\selectfont\catcode`\^=\active\def^{\ifmmode\sp\else\^{}\fi}\catcode`\%=\active\def
\end{pgfscope}%
\begin{pgfscope}%
\pgfsetbuttcap%
\pgfsetroundjoin%
\definecolor{currentfill}{rgb}{0.000000,0.000000,0.000000}%
\pgfsetfillcolor{currentfill}%
\pgfsetlinewidth{0.803000pt}%
\definecolor{currentstroke}{rgb}{0.000000,0.000000,0.000000}%
\pgfsetstrokecolor{currentstroke}%
\pgfsetdash{}{0pt}%
\pgfsys@defobject{currentmarker}{\pgfqpoint{-0.048611in}{0.000000in}}{\pgfqpoint{-0.000000in}{0.000000in}}{%
\pgfpathmoveto{\pgfqpoint{-0.000000in}{0.000000in}}%
\pgfpathlineto{\pgfqpoint{-0.048611in}{0.000000in}}%
\pgfusepath{stroke,fill}%
}%
\begin{pgfscope}%
\pgfsys@transformshift{0.938111in}{4.368111in}%
\pgfsys@useobject{currentmarker}{}%
\end{pgfscope}%
\end{pgfscope}%
\begin{pgfscope}%
\definecolor{textcolor}{rgb}{0.000000,0.000000,0.000000}%
\pgfsetstrokecolor{textcolor}%
\pgfsetfillcolor{textcolor}%
\pgftext[x=0.377333in, y=4.290999in, left, base]{\color{textcolor}{\fontfamily{\familydefault}\fontsize{16.000000}{19.200000}\selectfont\catcode`\^=\active\def^{\ifmmode\sp\else\^{}\fi}\catcode`\%=\active\def
\end{pgfscope}%
\begin{pgfscope}%
\definecolor{textcolor}{rgb}{0.000000,0.000000,0.000000}%
\pgfsetstrokecolor{textcolor}%
\pgfsetfillcolor{textcolor}%
\pgftext[x=0.321778in,y=2.520111in,,bottom,rotate=90.000000]{\color{textcolor}{\fontfamily{\familydefault}\fontsize{16.000000}{19.200000}\selectfont\catcode`\^=\active\def^{\ifmmode\sp\else\^{}\fi}\catcode`\%=\active\def
\end{pgfscope}%
\begin{pgfscope}%
\pgfpathrectangle{\pgfqpoint{0.938111in}{0.672111in}}{\pgfqpoint{4.960000in}{3.696000in}}%
\pgfusepath{clip}%
\pgfsetrectcap%
\pgfsetroundjoin%
\pgfsetlinewidth{1.505625pt}%
\definecolor{currentstroke}{rgb}{1.000000,0.000000,0.000000}%
\pgfsetstrokecolor{currentstroke}%
\pgfsetdash{}{0pt}%
\pgfpathmoveto{\pgfqpoint{0.934778in}{2.099118in}}%
\pgfpathlineto{\pgfqpoint{0.938111in}{2.109460in}}%
\pgfpathlineto{\pgfqpoint{0.969566in}{2.011868in}}%
\pgfpathlineto{\pgfqpoint{1.001022in}{1.757031in}}%
\pgfpathlineto{\pgfqpoint{1.032477in}{1.435277in}}%
\pgfpathlineto{\pgfqpoint{1.063933in}{1.138352in}}%
\pgfpathlineto{\pgfqpoint{1.095388in}{0.919456in}}%
\pgfpathlineto{\pgfqpoint{1.126844in}{0.786063in}}%
\pgfpathlineto{\pgfqpoint{1.158299in}{0.717725in}}%
\pgfpathlineto{\pgfqpoint{1.189755in}{0.687987in}}%
\pgfpathlineto{\pgfqpoint{1.221210in}{0.676901in}}%
\pgfpathlineto{\pgfqpoint{1.252666in}{0.673337in}}%
\pgfpathlineto{\pgfqpoint{1.284121in}{0.672389in}}%
\pgfpathlineto{\pgfqpoint{1.409943in}{0.672328in}}%
\pgfpathlineto{\pgfqpoint{1.881775in}{0.672199in}}%
\pgfpathlineto{\pgfqpoint{2.039053in}{0.672476in}}%
\pgfpathlineto{\pgfqpoint{2.070508in}{0.673182in}}%
\pgfpathlineto{\pgfqpoint{2.101963in}{0.675367in}}%
\pgfpathlineto{\pgfqpoint{2.133419in}{0.680699in}}%
\pgfpathlineto{\pgfqpoint{2.164874in}{0.690569in}}%
\pgfpathlineto{\pgfqpoint{2.196330in}{0.704635in}}%
\pgfpathlineto{\pgfqpoint{2.227785in}{0.737661in}}%
\pgfpathlineto{\pgfqpoint{2.259241in}{0.863665in}}%
\pgfpathlineto{\pgfqpoint{2.290696in}{1.176978in}}%
\pgfpathlineto{\pgfqpoint{2.322152in}{1.739861in}}%
\pgfpathlineto{\pgfqpoint{2.353607in}{2.516285in}}%
\pgfpathlineto{\pgfqpoint{2.385063in}{3.321688in}}%
\pgfpathlineto{\pgfqpoint{2.416518in}{3.873682in}}%
\pgfpathlineto{\pgfqpoint{2.447974in}{3.945481in}}%
\pgfpathlineto{\pgfqpoint{2.479429in}{3.513178in}}%
\pgfpathlineto{\pgfqpoint{2.510885in}{2.769328in}}%
\pgfpathlineto{\pgfqpoint{2.542340in}{1.991424in}}%
\pgfpathlineto{\pgfqpoint{2.573796in}{1.382573in}}%
\pgfpathlineto{\pgfqpoint{2.605251in}{1.003487in}}%
\pgfpathlineto{\pgfqpoint{2.636707in}{0.809718in}}%
\pgfpathlineto{\pgfqpoint{2.668162in}{0.725691in}}%
\pgfpathlineto{\pgfqpoint{2.699618in}{0.693196in}}%
\pgfpathlineto{\pgfqpoint{2.731073in}{0.681435in}}%
\pgfpathlineto{\pgfqpoint{2.762528in}{0.677361in}}%
\pgfpathlineto{\pgfqpoint{2.793984in}{0.676070in}}%
\pgfpathlineto{\pgfqpoint{2.825439in}{0.676141in}}%
\pgfpathlineto{\pgfqpoint{2.888350in}{0.677116in}}%
\pgfpathlineto{\pgfqpoint{2.919806in}{0.676704in}}%
\pgfpathlineto{\pgfqpoint{3.014172in}{0.673285in}}%
\pgfpathlineto{\pgfqpoint{3.077083in}{0.674715in}}%
\pgfpathlineto{\pgfqpoint{3.139994in}{0.675950in}}%
\pgfpathlineto{\pgfqpoint{3.202905in}{0.676249in}}%
\pgfpathlineto{\pgfqpoint{3.265816in}{0.675440in}}%
\pgfpathlineto{\pgfqpoint{3.328727in}{0.674213in}}%
\pgfpathlineto{\pgfqpoint{3.360183in}{0.674014in}}%
\pgfpathlineto{\pgfqpoint{3.486004in}{0.675085in}}%
\pgfpathlineto{\pgfqpoint{3.548915in}{0.674398in}}%
\pgfpathlineto{\pgfqpoint{3.611826in}{0.673655in}}%
\pgfpathlineto{\pgfqpoint{3.643282in}{0.674003in}}%
\pgfpathlineto{\pgfqpoint{3.674737in}{0.675337in}}%
\pgfpathlineto{\pgfqpoint{3.706193in}{0.678005in}}%
\pgfpathlineto{\pgfqpoint{3.737648in}{0.682073in}}%
\pgfpathlineto{\pgfqpoint{3.769104in}{0.687190in}}%
\pgfpathlineto{\pgfqpoint{3.832015in}{0.698090in}}%
\pgfpathlineto{\pgfqpoint{3.863470in}{0.702978in}}%
\pgfpathlineto{\pgfqpoint{3.894926in}{0.706968in}}%
\pgfpathlineto{\pgfqpoint{3.926381in}{0.709100in}}%
\pgfpathlineto{\pgfqpoint{3.957837in}{0.708117in}}%
\pgfpathlineto{\pgfqpoint{3.989292in}{0.703491in}}%
\pgfpathlineto{\pgfqpoint{4.020748in}{0.696182in}}%
\pgfpathlineto{\pgfqpoint{4.052203in}{0.688220in}}%
\pgfpathlineto{\pgfqpoint{4.083659in}{0.681509in}}%
\pgfpathlineto{\pgfqpoint{4.115114in}{0.676926in}}%
\pgfpathlineto{\pgfqpoint{4.146569in}{0.674307in}}%
\pgfpathlineto{\pgfqpoint{4.178025in}{0.672999in}}%
\pgfpathlineto{\pgfqpoint{4.209480in}{0.672406in}}%
\pgfpathlineto{\pgfqpoint{4.303847in}{0.672185in}}%
\pgfpathlineto{\pgfqpoint{4.398213in}{0.672579in}}%
\pgfpathlineto{\pgfqpoint{4.492580in}{0.672934in}}%
\pgfpathlineto{\pgfqpoint{4.838590in}{0.672434in}}%
\pgfpathlineto{\pgfqpoint{4.995867in}{0.672771in}}%
\pgfpathlineto{\pgfqpoint{5.216056in}{0.672496in}}%
\pgfpathlineto{\pgfqpoint{5.404789in}{0.672931in}}%
\pgfpathlineto{\pgfqpoint{5.719343in}{0.672391in}}%
\pgfpathlineto{\pgfqpoint{5.901444in}{0.672405in}}%
\pgfpathlineto{\pgfqpoint{5.901444in}{0.672405in}}%
\pgfusepath{stroke}%
\end{pgfscope}%
\begin{pgfscope}%
\pgfpathrectangle{\pgfqpoint{0.938111in}{0.672111in}}{\pgfqpoint{4.960000in}{3.696000in}}%
\pgfusepath{clip}%
\pgfsetbuttcap%
\pgfsetroundjoin%
\pgfsetlinewidth{1.505625pt}%
\definecolor{currentstroke}{rgb}{0.000000,0.500000,0.000000}%
\pgfsetstrokecolor{currentstroke}%
\pgfsetdash{{5.550000pt}{2.400000pt}}{0.000000pt}%
\pgfpathmoveto{\pgfqpoint{0.934778in}{2.099155in}}%
\pgfpathlineto{\pgfqpoint{0.938111in}{2.109497in}}%
\pgfpathlineto{\pgfqpoint{0.969565in}{2.011910in}}%
\pgfpathlineto{\pgfqpoint{1.001020in}{1.757083in}}%
\pgfpathlineto{\pgfqpoint{1.032474in}{1.435338in}}%
\pgfpathlineto{\pgfqpoint{1.063929in}{1.138413in}}%
\pgfpathlineto{\pgfqpoint{1.095383in}{0.919509in}}%
\pgfpathlineto{\pgfqpoint{1.126838in}{0.786104in}}%
\pgfpathlineto{\pgfqpoint{1.158292in}{0.717752in}}%
\pgfpathlineto{\pgfqpoint{1.189747in}{0.688002in}}%
\pgfpathlineto{\pgfqpoint{1.221201in}{0.676903in}}%
\pgfpathlineto{\pgfqpoint{1.252656in}{0.673328in}}%
\pgfpathlineto{\pgfqpoint{1.284110in}{0.672371in}}%
\pgfpathlineto{\pgfqpoint{1.472838in}{0.672220in}}%
\pgfpathlineto{\pgfqpoint{1.755928in}{0.672160in}}%
\pgfpathlineto{\pgfqpoint{2.039019in}{0.672473in}}%
\pgfpathlineto{\pgfqpoint{2.070473in}{0.673178in}}%
\pgfpathlineto{\pgfqpoint{2.101928in}{0.675360in}}%
\pgfpathlineto{\pgfqpoint{2.133382in}{0.680684in}}%
\pgfpathlineto{\pgfqpoint{2.164837in}{0.690541in}}%
\pgfpathlineto{\pgfqpoint{2.196291in}{0.704588in}}%
\pgfpathlineto{\pgfqpoint{2.227746in}{0.737570in}}%
\pgfpathlineto{\pgfqpoint{2.259200in}{0.863461in}}%
\pgfpathlineto{\pgfqpoint{2.290655in}{1.176557in}}%
\pgfpathlineto{\pgfqpoint{2.322109in}{1.739172in}}%
\pgfpathlineto{\pgfqpoint{2.353564in}{2.515422in}}%
\pgfpathlineto{\pgfqpoint{2.385018in}{3.320926in}}%
\pgfpathlineto{\pgfqpoint{2.416473in}{3.873362in}}%
\pgfpathlineto{\pgfqpoint{2.447927in}{3.945794in}}%
\pgfpathlineto{\pgfqpoint{2.479382in}{3.514034in}}%
\pgfpathlineto{\pgfqpoint{2.510836in}{2.770411in}}%
\pgfpathlineto{\pgfqpoint{2.542291in}{1.992401in}}%
\pgfpathlineto{\pgfqpoint{2.573745in}{1.383265in}}%
\pgfpathlineto{\pgfqpoint{2.605200in}{1.003886in}}%
\pgfpathlineto{\pgfqpoint{2.636654in}{0.809913in}}%
\pgfpathlineto{\pgfqpoint{2.668109in}{0.725774in}}%
\pgfpathlineto{\pgfqpoint{2.699564in}{0.693227in}}%
\pgfpathlineto{\pgfqpoint{2.731018in}{0.681441in}}%
\pgfpathlineto{\pgfqpoint{2.762473in}{0.677357in}}%
\pgfpathlineto{\pgfqpoint{2.793927in}{0.676059in}}%
\pgfpathlineto{\pgfqpoint{2.825382in}{0.676123in}}%
\pgfpathlineto{\pgfqpoint{2.888291in}{0.677099in}}%
\pgfpathlineto{\pgfqpoint{2.919745in}{0.676692in}}%
\pgfpathlineto{\pgfqpoint{3.014109in}{0.673267in}}%
\pgfpathlineto{\pgfqpoint{3.077018in}{0.674701in}}%
\pgfpathlineto{\pgfqpoint{3.139927in}{0.675934in}}%
\pgfpathlineto{\pgfqpoint{3.202836in}{0.676239in}}%
\pgfpathlineto{\pgfqpoint{3.265745in}{0.675448in}}%
\pgfpathlineto{\pgfqpoint{3.328654in}{0.674237in}}%
\pgfpathlineto{\pgfqpoint{3.360108in}{0.674030in}}%
\pgfpathlineto{\pgfqpoint{3.485926in}{0.675054in}}%
\pgfpathlineto{\pgfqpoint{3.548835in}{0.674380in}}%
\pgfpathlineto{\pgfqpoint{3.611744in}{0.673649in}}%
\pgfpathlineto{\pgfqpoint{3.643199in}{0.673998in}}%
\pgfpathlineto{\pgfqpoint{3.674653in}{0.675331in}}%
\pgfpathlineto{\pgfqpoint{3.706108in}{0.677995in}}%
\pgfpathlineto{\pgfqpoint{3.737562in}{0.682061in}}%
\pgfpathlineto{\pgfqpoint{3.769017in}{0.687176in}}%
\pgfpathlineto{\pgfqpoint{3.831926in}{0.698080in}}%
\pgfpathlineto{\pgfqpoint{3.863380in}{0.702970in}}%
\pgfpathlineto{\pgfqpoint{3.894835in}{0.706964in}}%
\pgfpathlineto{\pgfqpoint{3.926289in}{0.709103in}}%
\pgfpathlineto{\pgfqpoint{3.957744in}{0.708128in}}%
\pgfpathlineto{\pgfqpoint{3.989198in}{0.703510in}}%
\pgfpathlineto{\pgfqpoint{4.020653in}{0.696206in}}%
\pgfpathlineto{\pgfqpoint{4.052108in}{0.688243in}}%
\pgfpathlineto{\pgfqpoint{4.083562in}{0.681525in}}%
\pgfpathlineto{\pgfqpoint{4.115017in}{0.676937in}}%
\pgfpathlineto{\pgfqpoint{4.146471in}{0.674313in}}%
\pgfpathlineto{\pgfqpoint{4.177926in}{0.673003in}}%
\pgfpathlineto{\pgfqpoint{4.209380in}{0.672411in}}%
\pgfpathlineto{\pgfqpoint{4.303744in}{0.672189in}}%
\pgfpathlineto{\pgfqpoint{4.429562in}{0.672769in}}%
\pgfpathlineto{\pgfqpoint{4.523925in}{0.672863in}}%
\pgfpathlineto{\pgfqpoint{4.618289in}{0.672578in}}%
\pgfpathlineto{\pgfqpoint{4.744107in}{0.672688in}}%
\pgfpathlineto{\pgfqpoint{4.838470in}{0.672435in}}%
\pgfpathlineto{\pgfqpoint{4.995743in}{0.672770in}}%
\pgfpathlineto{\pgfqpoint{5.215924in}{0.672501in}}%
\pgfpathlineto{\pgfqpoint{5.404652in}{0.672935in}}%
\pgfpathlineto{\pgfqpoint{5.750651in}{0.672361in}}%
\pgfpathlineto{\pgfqpoint{5.901444in}{0.672411in}}%
\pgfpathlineto{\pgfqpoint{5.901444in}{0.672411in}}%
\pgfusepath{stroke}%
\end{pgfscope}%
\begin{pgfscope}%
\pgfsetrectcap%
\pgfsetmiterjoin%
\pgfsetlinewidth{1.003750pt}%
\definecolor{currentstroke}{rgb}{0.000000,0.000000,0.000000}%
\pgfsetstrokecolor{currentstroke}%
\pgfsetdash{}{0pt}%
\pgfpathmoveto{\pgfqpoint{0.938111in}{0.672111in}}%
\pgfpathlineto{\pgfqpoint{0.938111in}{4.368111in}}%
\pgfusepath{stroke}%
\end{pgfscope}%
\begin{pgfscope}%
\pgfsetrectcap%
\pgfsetmiterjoin%
\pgfsetlinewidth{1.003750pt}%
\definecolor{currentstroke}{rgb}{0.000000,0.000000,0.000000}%
\pgfsetstrokecolor{currentstroke}%
\pgfsetdash{}{0pt}%
\pgfpathmoveto{\pgfqpoint{5.898111in}{0.672111in}}%
\pgfpathlineto{\pgfqpoint{5.898111in}{4.368111in}}%
\pgfusepath{stroke}%
\end{pgfscope}%
\begin{pgfscope}%
\pgfsetrectcap%
\pgfsetmiterjoin%
\pgfsetlinewidth{1.003750pt}%
\definecolor{currentstroke}{rgb}{0.000000,0.000000,0.000000}%
\pgfsetstrokecolor{currentstroke}%
\pgfsetdash{}{0pt}%
\pgfpathmoveto{\pgfqpoint{0.938111in}{0.672111in}}%
\pgfpathlineto{\pgfqpoint{5.898111in}{0.672111in}}%
\pgfusepath{stroke}%
\end{pgfscope}%
\begin{pgfscope}%
\pgfsetrectcap%
\pgfsetmiterjoin%
\pgfsetlinewidth{1.003750pt}%
\definecolor{currentstroke}{rgb}{0.000000,0.000000,0.000000}%
\pgfsetstrokecolor{currentstroke}%
\pgfsetdash{}{0pt}%
\pgfpathmoveto{\pgfqpoint{0.938111in}{4.368111in}}%
\pgfpathlineto{\pgfqpoint{5.898111in}{4.368111in}}%
\pgfusepath{stroke}%
\end{pgfscope}%
\begin{pgfscope}%
\pgfsetbuttcap%
\pgfsetmiterjoin%
\definecolor{currentfill}{rgb}{1.000000,1.000000,1.000000}%
\pgfsetfillcolor{currentfill}%
\pgfsetfillopacity{0.800000}%
\pgfsetlinewidth{1.003750pt}%
\definecolor{currentstroke}{rgb}{0.800000,0.800000,0.800000}%
\pgfsetstrokecolor{currentstroke}%
\pgfsetstrokeopacity{0.800000}%
\pgfsetdash{}{0pt}%
\pgfpathmoveto{\pgfqpoint{3.984079in}{3.570333in}}%
\pgfpathlineto{\pgfqpoint{5.742555in}{3.570333in}}%
\pgfpathquadraticcurveto{\pgfqpoint{5.787000in}{3.570333in}}{\pgfqpoint{5.787000in}{3.614778in}}%
\pgfpathlineto{\pgfqpoint{5.787000in}{4.212555in}}%
\pgfpathquadraticcurveto{\pgfqpoint{5.787000in}{4.256999in}}{\pgfqpoint{5.742555in}{4.256999in}}%
\pgfpathlineto{\pgfqpoint{3.984079in}{4.256999in}}%
\pgfpathquadraticcurveto{\pgfqpoint{3.939634in}{4.256999in}}{\pgfqpoint{3.939634in}{4.212555in}}%
\pgfpathlineto{\pgfqpoint{3.939634in}{3.614778in}}%
\pgfpathquadraticcurveto{\pgfqpoint{3.939634in}{3.570333in}}{\pgfqpoint{3.984079in}{3.570333in}}%
\pgfpathlineto{\pgfqpoint{3.984079in}{3.570333in}}%
\pgfpathclose%
\pgfusepath{stroke,fill}%
\end{pgfscope}%
\begin{pgfscope}%
\pgfsetrectcap%
\pgfsetroundjoin%
\pgfsetlinewidth{1.505625pt}%
\definecolor{currentstroke}{rgb}{1.000000,0.000000,0.000000}%
\pgfsetstrokecolor{currentstroke}%
\pgfsetdash{}{0pt}%
\pgfpathmoveto{\pgfqpoint{4.028523in}{4.090333in}}%
\pgfpathlineto{\pgfqpoint{4.250746in}{4.090333in}}%
\pgfpathlineto{\pgfqpoint{4.472968in}{4.090333in}}%
\pgfusepath{stroke}%
\end{pgfscope}%
\begin{pgfscope}%
\definecolor{textcolor}{rgb}{0.000000,0.000000,0.000000}%
\pgfsetstrokecolor{textcolor}%
\pgfsetfillcolor{textcolor}%
\pgftext[x=4.650746in,y=4.012555in,left,base]{\color{textcolor}{\fontfamily{\familydefault}\fontsize{16.000000}{19.200000}\selectfont\catcode`\^=\active\def^{\ifmmode\sp\else\^{}\fi}\catcode`\%=\active\def
\end{pgfscope}%
\begin{pgfscope}%
\pgfsetbuttcap%
\pgfsetroundjoin%
\pgfsetlinewidth{1.505625pt}%
\definecolor{currentstroke}{rgb}{0.000000,0.500000,0.000000}%
\pgfsetstrokecolor{currentstroke}%
\pgfsetdash{{5.550000pt}{2.400000pt}}{0.000000pt}%
\pgfpathmoveto{\pgfqpoint{4.028523in}{3.780333in}}%
\pgfpathlineto{\pgfqpoint{4.250746in}{3.780333in}}%
\pgfpathlineto{\pgfqpoint{4.472968in}{3.780333in}}%
\pgfusepath{stroke}%
\end{pgfscope}%
\begin{pgfscope}%
\definecolor{textcolor}{rgb}{0.000000,0.000000,0.000000}%
\pgfsetstrokecolor{textcolor}%
\pgfsetfillcolor{textcolor}%
\pgftext[x=4.650746in,y=3.702555in,left,base]{\color{textcolor}{\fontfamily{\familydefault}\fontsize{16.000000}{19.200000}\selectfont\catcode`\^=\active\def^{\ifmmode\sp\else\^{}\fi}\catcode`\%=\active\def
\end{pgfscope}%
\end{pgfpicture}%
\makeatother%
\endgroup%